\documentclass[oneside, a4paper, onecolumn, 11pt]{article}
\usepackage[left=2.5cm,top=2.5cm,bottom=2.5cm,right=2.5cm]{geometry}
\usepackage{amsmath}
\usepackage{graphicx}
\usepackage{fancyhdr}
\usepackage{calc}
\usepackage{amssymb}
\usepackage[super,sort&compress,comma]{natbib}
\usepackage{setspace}
\usepackage{amsfonts}
\usepackage{commath}
\usepackage[innercaption]{sidecap}
\usepackage[framemethod=TikZ]{mdframed}
\usepackage{wrapfig}
\usepackage{hyperref}
\usepackage{mathrsfs}
\usepackage{lineno}

\newcommand{\N}{\mathcal{N}}        
\newcommand{\ra}[1]{r_{#1}}                 

\newcommand{\Plxy}[2]{P_{R_\pm}\left(x,y; t = 0\right)}   
\newcommand{\Q}{q}

\newcommand{\lmin}{\lambda_{\rm min}}      
\newcommand{\cumtwo}[4]{K_{#1,#2}\left( #3,#4 \right) }
\newcommand{\E}{{\rm E}} 
\newcommand{\V}{\mathcal{V}} 
\newcommand{\Norm}{\mathbb{N}} 
 
\newcommand{\av}[1]{\langle #1 \rangle}

\expandafter\def\expandafter\normalsize\expandafter{%
	\normalsize
	\setlength\abovedisplayskip{0pt}
	\setlength\belowdisplayskip{5pt}
	\setlength\abovedisplayshortskip{0pt}
	\setlength\belowdisplayshortskip{5pt}
}

\definecolor{Gray}{gray}{0.75}

\newmdenv[backgroundcolor=Gray, leftmargin = 0pt, rightmargin = 0pt, linewidth = 0pt, roundcorner = 2 pt, innerleftmargin=5pt, innerrightmargin=5pt, innertopmargin=5pt, innerbottommargin=5pt]{Frame}

\begin{document}

\title{\LARGE \color {blue} \textbf{Efficient treatment of heterogeneous} \\ \textbf{malignant cell populations}}
\author{Uzi Harush,$^{1,2}$ Ravid Straussman$^3$ and Baruch Barzel$^{1,2,4}$}

\maketitle

\begin{enumerate}
\footnotesize
\item
Department of Mathematics, Bar-Ilan University, Ramat-Gan, Israel 52900.
\item
The Gonda Multidisciplinary Brain Research Center, Bar-Ilan University, Ramat-Gan, Israel 52900.
\item
Department of Molecular Cell Biology, Weizmann Institute of Science, Rehovot, Israel 76100.
\item
The Network Science Institute, Northeastern University, Boston, MA., 02115, US. 
\end{enumerate}
\vspace{4mm}

\textbf{When confronted with an undesired cell population, such as bacterial infections or tumors, we seek the most effective treatment, designed to eliminate the population as rapidly as possible. A common practice is to monitor the cells' short-term response to the treatment, and from that, extrapolate the eventual treatment outcome, \textit{i.e}.\ will it eradicate the cells, and if yes - at what timescales. Underlying this approach is the assumption that the cells exhibit a homogeneous response to the treatment, and hence the early response patterns can be naturally extended to later times. Recent experiments on cancer cell populations, however, indicate a significant level of cellular heterogeneity, undermining this classic assessment protocol of treatment efficacy. We, therefore, develop here a stochastic framework, to analytically predict the temporal dynamics of a heterogeneous cell population. Quite often, we find, the average cellular parameters, governing the short-term response, fail to predict the actual treatment outcome. In contrast, our analysis, which also incorporates the population's variability, helps identify the relevant statistical parameters, which in turn, enable us to predict the full trajectory of the cell population, and specifically - the likelihood and typical timescales for remission.  
}

The exponential decay is a hallmark of many population dynamical systems driven by birth and death processes. \cite{Susman2018,Brenner2015,Tamari2016,Charlebois2019,Di2008,Giorno2020,Bacaer2011} Some examples include bacterial populations under antibiotic treatment, \cite{Keren2004,Balaban2004,Balaban2019,Andersson2003,Bakkeren2020,Shomar2022,Brauner2016,Bigger1944} chemical compounds undergoing degradation, \cite{Barzel2011,Barzel2012} or radioactive processes\cite{Bobkova2020}, all of whom follow the universal form 
\begin{equation}
N_t = N_0 e^{-\lambda t}.
\label{ExponentialDecay}
\end{equation}
Under $\lambda > 0$ this captures a rapid exponential decay of the population, with a typical timescale $\tau = 1/\lambda$, ensuring the population's elimination within $t$ sufficiently greater than $\tau$. 

In clinical applications this exponential nature of $N_t$ is often taken for granted, \cite{Bartolomeo2021,Johnson2019} \textit{e.g}., when estimating the desired dose and duration of a specific treatment. For example, consider antibiotics administered against a bacterial infection. By tracking the response of the bacterial population over the course of several hours, one can evaluate $\lambda$, extrapolate the long term behavior of $N_t$ via (\ref{ExponentialDecay}), and by that, derive the treatment duration $T$ that ensures $N_{t = T} \ll 1$, \textit{i.e}.\ the time when \textit{almost} no surviving bacteria remain. \cite{Salem2006} Hence a short-term observation of $N_t$ can directly provide us with information on the treatment effectiveness over time. Specifically, setting $N_T \ll 1$ in Eq.\ (\ref{ExponentialDecay}) we obtain
\begin{equation}
T \gtrsim \dfrac{1}{\lambda} \ln N_0,
\label{TLogN0}
\end{equation}  
a logarithmic dependence on the initial population size, ensuring that even a macroscopic population, with $N_0 \to \infty$, can be eradicated within a relatively small (logarithmically bounded) timescale.

\begin{figure}[h!]
\centering
\includegraphics[width=0.75\textwidth]{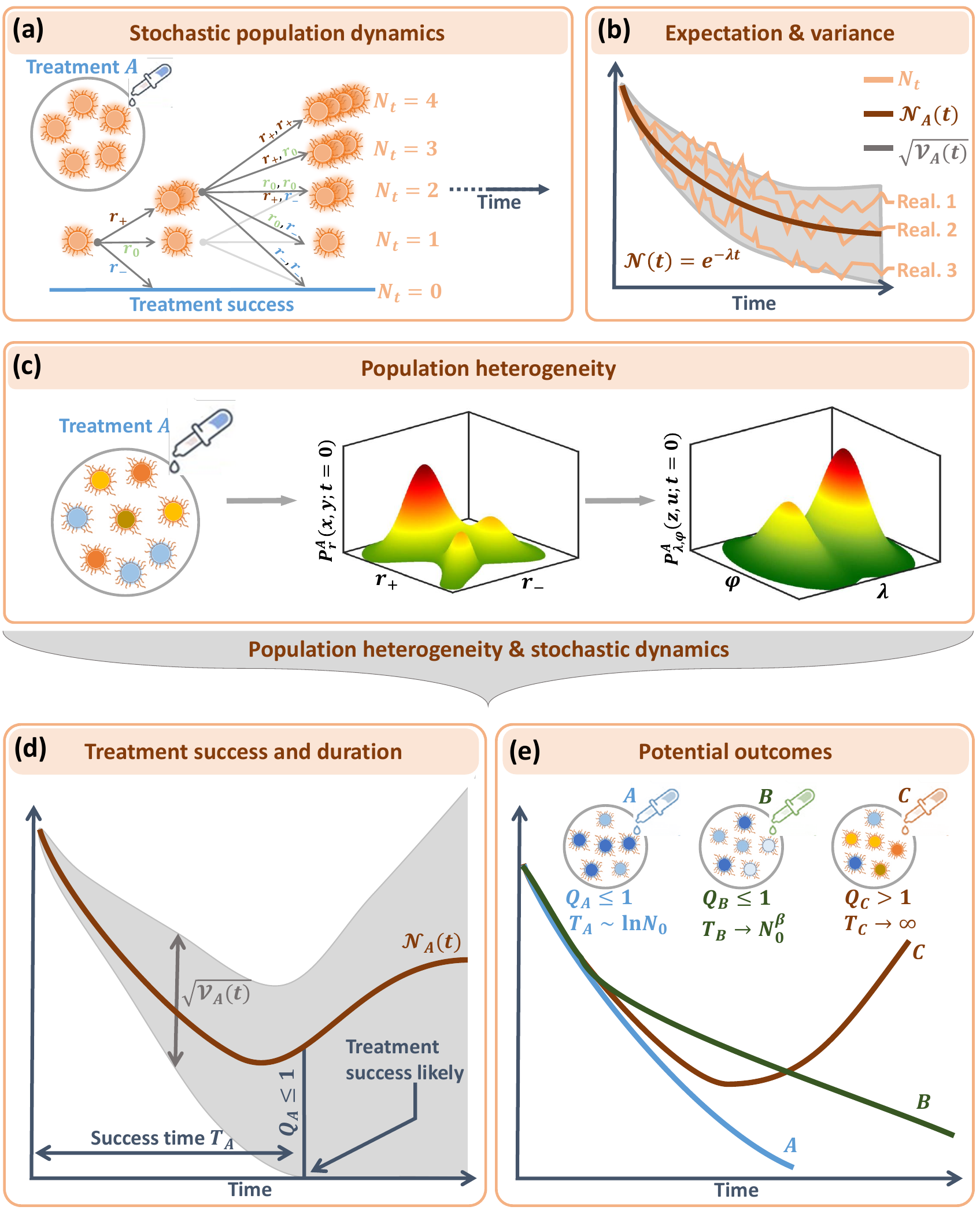}
\vspace{-3mm}
\caption{\footnotesize \textbf{\color{blue} The malignancy-treatment interplay}.\
(a) The dynamics of a homogeneous cell population under treatment $A$. In each time step cells duplicate, remain idle or decay stochastically at rates $\ra+,\ra0$ and $\ra-$, respectively. These rates depend on the cell-type and on $A$'s efficacy. The resulting population dynamics is characterized by the discrete random variable $N_t$, which describes the cell population at time $t$. If at some instance $N_t$ hits zero (blue horizontal line), the malignant population is eliminated and the treatment is deemed successful.
(b) Due to the stochastic nature of $N_t$ each realization of the dynamics leads to a different outcome (orange lines). We can characterize this variability by reducing it into two statistical parameters:\ The \textit{expectation} $\N_A(t)$ (red), which decays exponentially at rate $\lambda = \ra- - \ra+$, and the \textit{standard deviation} $\sqrt{\V_A(t)}$ (grey), which captures the potential deviations from $\N_A(t)$. Most realizations fall within the boundaries of $\N_A(t) \pm \sqrt{\V_A(t)}$, \textit{i.e}.\ the shaded area.
(c) In reality, most treatments do not invoke a homogeneous response, but rather each cell is characterized by its idiosyncratic $\ra-,\ra0,\ra+$ (cell colors). Therefore, treatment $A$ is characterized by $P_r^A(x,y;t = 0)$, capturing the probability density that a random cell under $A$ has $\ra- \in (x,x + \dif x)$ and $\ra+ \in (y,y + \dif y)$. Our derivations indicate that the system can more naturally be described via the subsequent density function $P_{\lambda,\varphi}^A(z,u;t = 0)$, where $\lambda = \ra- - \ra+$ and $\varphi = \ra- + \ra+$. Our goal is to characterized the population dynamics under the combination of both population heterogeneity (panel c) and the stochastic birth/death sequences of each specific cell lineage (panel a).   
(d) Our analysis predicts $\N_A(t)$ (red) and $\V_A(t)$ (grey) as obtained from the heterogeneous and stochastic dynamics of panels (a) and (c). This allows us to evaluate the bounds of the observed treatment dynamics $N_t$ to be within the grey shaded area, offering two crucial statistics:\ $Q_A$ helps predict whether the shaded area crosses the zero population line. When such crossing occurs ($Q_A \le 1$) it is likely that the random variable $N_t$ will, at some point, hit zero, \textit{i.e}.\ \textit{treatment success}. $T_A$ captures the typical timescale to reach $Q_A \le 1$, hence predicting the \textit{treatment efficiency}. We seek treatments with $Q_A \le 1$ and $T_A$ small.
(e) Extracting our two statistics $Q_A,T_A$ from the initial population parameters ($N_0$ and  $P_{\lambda,\varphi}^A(z,u;t = 0)$), we can predict the potential treatment outcomes:\ Exponential remission ($A$, blue), in which elimination is expected within logarithmic timescales (successful and efficient); Slow remission ($B$, green) where $T_B$ diverges polynomially with $N_0$ (successful bot not efficient); Recurrent ($C$, red), under which $Q_A > 1$, and hence the population is unlikely to ever reach $N_t = 0$, and most chances are that it will reemerge (not successful nor efficient).  
}
\label{Fig1}
\end{figure}

\begin{figure}[h!]
\centering
\includegraphics[width=0.9\textwidth]{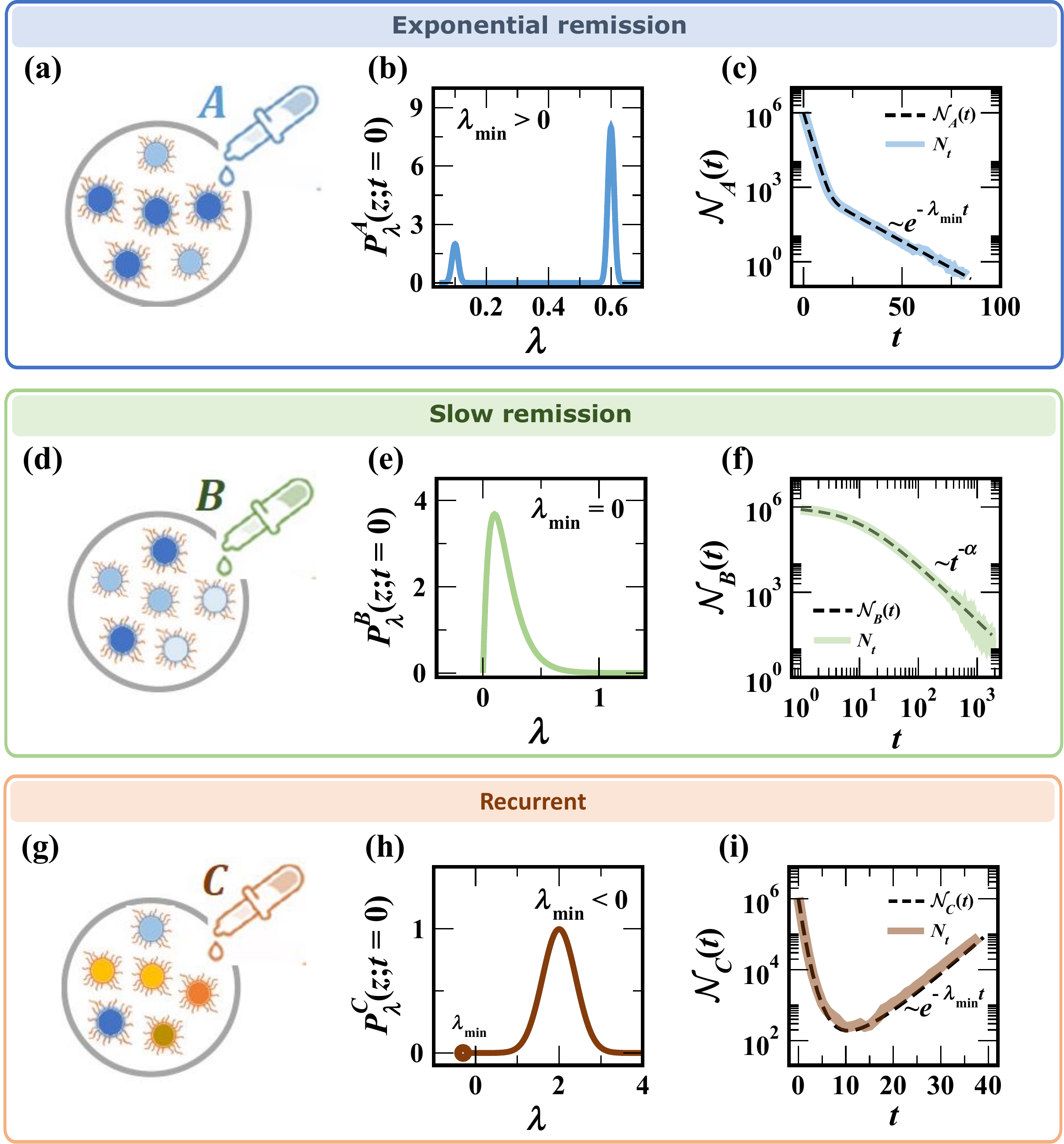}
\vspace{-3mm}
\caption{\footnotesize \textbf{\color{blue} Three classes of the \textit{expected} population dynamics $\N_A(t)$}.\
Under $P^A_{\lambda}(z;t = 0)$ in (\ref{Hahn}) we predict that $\N_A(t)$ can follow three distinct paths.
(a) In case $\lmin > 0$ we observe exponential remission. 
(b) As an example we consider a bi-modal $P^A_{\lambda}(z;t = 0)$, as commonly observed in antibiotic bacterial persistence:\ most bacteria exhibit a rapid decay (large $\lambda$), while a small minority persist (small $\lambda$). These two $\lambda$ values are depicted by dark/light blue shades in panel (a). 
(c) The result is that $\N_A(t) \sim e^{-\lmin t}$ (dashed line), driven by the slowest decaying sub-population. Still, despite the slow $\lmin$, this class predicts a long-term exponential decay of the form of Eq.\ (\ref{ExponentialDecay}). Such dynamics guarantees to eliminate the undesired population within logarithmic timescales, as in (\ref{TLogN0}). We also show an actual stochastic realization of $N_t$ (blue) conducted under the bi-modal $P_\lambda^A(z;t = 0)$ of panel (b).  
(d) - (f) In case $P^A_{\lambda}(z;t = 0)$ is bounded at $z = 0$ we have $\lmin = 0$, leading to slow remission. Such treatments completely suppress growth (\textit{i.e}.\ no $\lambda < 0$), but allow for arbitrarily slow decay rates (blue shades). Here $\N_A(t)$ (dashed line) follows a slow polynomial death ($\sim t^{-\alpha}$), and hence the typical treatment duration diverges with $N_0$, resulting in \textit{successful} but \textit{inefficient} treatment. Such slow decay is also observed under a stochastic realization (green) conducted with $P_\lambda^A(z;t = 0)$ of panel (e). 
(g) - (i) In the presence of negative $\lmin$ (red cells), \textit{e.g}.\ under normally distributed rates, the expectation will follow a recurrent dynamics, reemerging with a long term exponential growth, driven by the small seed of positively growing cells that may exist within the original population (panel (i), $\N_A(t)$, dashed line; $N_t$, red). Note that here, despite taking $P_\lambda^A(z;t = 0)$ from a normal distribution, which in principal allows $\lambda \in (-\infty,\infty)$, we still observe a finite $\lmin$ (panel (h), red dot), due to the finite initial population size $N_0$.
These classes, we emphasize, capture the \textit{expectation} $\N_A(t)$ but disregard potential stochastic deviations, \textit{i.e}.\ $\V_A(t)$, as represented by the grey shaded in Fig.\ \ref{Fig1}b,d. Taking the latter into account, via $Q_A$ and $T_A$ in (\ref{Qmin}) and (\ref{T}), we detect conditions where, \textit{e.g}., extinction may precede recurrence. Under such conditions, while $\N_A(t)$ reemerges, as per the recurrent class, the actual observed treatment outcome ($N_t$) may, with high likelihood, be successful and, in fact, reach $N_t = 0$, thus averting the recurrent branch. Therefore, our full analysis does not stop at $\N_A(t)$, but also incorporates $\V_A(t)$ and its subsequent parameters $T_A$ and $Q_A$. 
}
\label{Fig2}
\end{figure}

The challenge is that many relevant systems incorporate a mixture of birth/death rates, as their initial population is intrinsically heterogeneous \cite{Kendall1948,Wang2010,Jacob2021} (Fig.\ \ref{Fig1}). A classic example for such mixture is observed in antibiotic persistence, \cite{Balaban2004,Bakkeren2020,Balaban2019,Andersson2003} where a majority of rapidly decaying bacteria (large $\lambda$) coexist alongside a small minority of persistent cells (small $\lambda$). This persistent minority outlives the designated treatment time $T$, leaving bacterial traces that may reemerge after the treatment is ceased. The resulting, well-documented, deviation from (\ref{ExponentialDecay}) can, in this case, be rectified quite simply by introducing two timescales:\ $\lambda_1$ and $\lambda_2$ for the rapid/slow decaying sub-populations, resulting in $N_t$ comprising a combination of two exponential functions (Fig.\ \ref{Fig2}a-c). 

More generally, however, such $\lambda$-heterogeneity, can follow a continuum of different birth/death rates, \cite{Jacob2021} resulting in a host of non-trivial deviations from Eq.\ (\ref{ExponentialDecay}). \cite{Amir2012} Our analysis finds that this diversity condenses into three potential classes of population dynamics (Fig.\ \ref{Fig2}):\ 
\textit{\color{blue} Exponential remission}, in which $N_t$ exhibits an effective rate $\lambda$, and hence the form (\ref{ExponentialDecay}) continues to drive the decay; \textit{\color{blue} Slow remission}, in which $N_t$ is trapped by a long power-law tail, thus lacking a bounded timescale for the population eradication; and finally, \textit{\color{blue} Recurrent}, where the initial decay is reversed, and the population reemerges after an initial period of decline. Therefore, to predict the clinical effectiveness of a suggested treatment $A$, we must go beyond the population average ($\lambda_A$), and account also for the $\lambda$-diversity in response to $A$. Namely, we wish to predict $N_t$ under a given density function $P^A_\lambda(z;t = 0)$. This function describes the probability to observe $\lambda \in (z, z + \dif z)$ within the initial ($t = 0$) cell population, under treatment $A$.   

Beyond the population heterogeneity $P^A_\lambda(z;t = 0)$, the observed response is also affected by the discrete and stochastic nature of $N_t$, which may advance along diverse trajectories,  \cite{Van1992} even if the conditions at $t = 0$ are fixed (Fig.\ \ref{Fig1}b, orange). For example, the recurrent dynamics, above, can be averted if, at some point $N_t$ becomes sufficiently small or fluctuating, and with high probability hits $N_t = 0$, \textit{i.e}.\ an extinction event from which there is no recovery. \cite{Ovaskainen2010,Dennehy2007} Therefore, here, we address precisely this question:\ given an initial cell population $N_0$, subject to treatment $A$, as characterized by $P^A_\lambda(z;t = 0)$, we seek the treatment efficacy, which boils down to two factors 
\textit{\color{blue} $\bullet$ Success rate}.\ Is $A$ likely to result in remission, \textit{i.e}.\ $N_t \to 0$ 
\textit{\color{blue} $\bullet$ Efficiency}.\ If yes, will it achieve such remission within a bounded timescale $T_A$ (Fig.\ \ref{Fig1}d,e).         

Quite generally, we find, that both questions can be answered through a small set of analytically tractable parameters, that can help us rank $A$'s potential effectiveness against the malignancy. Along the way, we encounter an array of clinically-relevant observations, such as a crossover effect, where a more effective treatment at small $t$, may, in the long-run, be superseded by a seemingly slower therapeutic; or a fat-tail phenomenon, \cite{Kaplan2021,Kauffman2003,Zaitsev2014} where an initially decaying population enters a state of slow polynomial death.

\vspace{5mm}
\textbf{\color{blue} \large Modeling heterogeneous population dynamics}

Consider a population of $N_0$ cells, representing \textit{e.g.}, a bacterial culture or a cancerous tumor. Under given environmental conditions, such as antibiotic treatment (bacteria) or chemotherapy (tumor), each cell exhibits an individual response, expressed via its replication and death rates, \cite{Crawford2014,Bressloff2014,Crawford2018} $\ra{+}$ and $\ra{-}$ (probability per unit time). This captures a \textit{continuous time stochastic process}, allowing us to track the population evolution via small (infinitesimal) time-steps $\dif t$ (Fig.\ \ref{Fig1}a). In each of these time-steps a cell my either replicate, with probability $\ra{+} \dif t$, die with probability $\ra{-} \dif t$, or remain idle, with probability $\ra0 \dif t = 1 - (\ra{+} + \ra{-}) \dif t$. Upon replication, an individual cell generates an identical offspring, that inherits its replication/death rates from its mother-cell. \cite{Jacob2021} Hence, in the present modeling framework, birth/death are stochastic, but the reproduction itself is mutation-free. \cite{Stewart2005,Zhang2022}

Under treatment, \textit{e.g}., chemotherapy, we seek to eradicate the undesired cells, by imposing conditions to drive all $\ra{-}$ up and $\ra{+}$ down, namely accelerate elimination and suppress growth. The cells, however, may exhibit a non-uniform response to our treatment - some being more persistent (higher $\ra{+}$ or lower $\ra{-}$) than others, and therefore the initial $N_0$ cell-population has distributed rates. Consequently, a given treatment $A$ is characterized not by a single decay rate $\lambda$, as appears in (\ref{ExponentialDecay}), but rather by a bi-variate probability density $P^A_r(x,y;t = 0)$, capturing the probability for a randomly selected cell at $t = 0$ to have, under treatment $A$, the birth/death rates $\ra{-} \in (x,x + \dif x)$ and $\ra{+} \in (y,y + \dif y)$; Fig.\ \ref{Fig1}c. 

Taken together, this framework introduces two sources of stochasticity:\ first, at $t = 0$, the birth/death rates of the original cell population are assigned probabilistically from $P^A_r(x,y;t = 0)$, characterizing $A$'s \textit{a priori} therapeutic efficacy. Then, once $A$ is administered and the dynamics ensue, each individual cell lineage grows/decays at random, driven by its assigned rates $\ra{-},\ra{+}$. The result is, that, under a given treatment $A$, each realization may evolve in time via a different trajectory, depending on the original $\ra{-},\ra{+}$ lottery, and on the subsequent birth/death sequence drawn by each cell. Hence, as opposed to the deterministic (\ref{ExponentialDecay}), here $N_t$ is a random variable, extracted from $P^A_t(N = n)$, \textit{i.e}.\ the probability that at time $t$ there remain $n$ cells in the population (Fig.\ \ref{Fig1}b, orange). 

If, as some point, we have $N_t = 0$, \textit{i.e}.\ extinction, the undesired cell population can no longer recover and the treatment is deemed successful. This, of course, cannot be guaranteed \textit{a priori}, given the random nature of $N_t$. Therefore, in this framework we resort to probabilistic success, namely we seek treatments that, with high likelihood, will lead to $N_t = 0$, \textit{i.e}.\ $P_t(N = 0) \lesssim 1$. Preferably this is reached at relatively small $t$, ensuring a quick remission. Hence, treatment $A$ is considered \textit{successful} if $N_t = 0$ is a probable outcome, for \textit{some} value of $t$; it is also deemed \textit{efficient}, if this value of $t$ is bounded.     

Predicting $P_t(N = n)$ is analytically prohibitive under the general assumptions presented above. However, as we show below, we can gain much insight by condensing the complete $N_t$-distribution into its two leading moments:\ the expectation $\N_A(t) \equiv \E_t(N) = \sum_{n = 0}^\infty n P^A_t(N = n)$, and the variance $\V_A(t) = \E_t(N^2) - \E_t^2(N)$. The former, $\N_A(t)$, describes the average population dynamics, as observed over $M \to \infty$ realizations (Fig.\ \ref{Fig1}d, red). It helps foresee the \textit{typical} response of the initial population to the treatment - namely, it captures treatment $A$'s \textit{expected trajectory}. The latter $\V_A(t)$, and its subsequent standard deviation $\sqrt{\V_A(t)}$, help quantify the potential diversity \textit{between} realizations, estimating the margins of uncertainty around $\N_A(t)$ (grey shaded). Hence, in an actual run, the observed $N_t$ (random variable) is likely to fall within 

\vspace{-1.5mm}
\begin{equation}
N_t \in 
\Big( \N_A(t) - \sqrt{\V_A(t)}, \N_A(t) + \sqrt{\V_A(t)} \Big),
\label{NtMargins}       
\end{equation}
namely - the expectation, $\N_A(t)$, plus or minus the likely fluctuation size, $\sqrt{\V_A(t)}$ 
{\color{blue} $\bullet$} 
We begin, below, by predicting $\N_A(t)$, then consider also the effect of $\V_A(t)$, to assess, via Eq.\ (\ref{NtMargins}), the relevant bounds on the system's observed stochastic behavior.

\vspace{4mm}
\textbf{\color{blue} \large The expected trajectory $\N_A(t)$}

Consider a single cell, with $\ra{-},\ra{+}$ extracted from $P^A_r(x,y;t = 0)$. On average, its lineage will evolve in time via Eq.\ (\ref{ExponentialDecay}) with $\lambda = \ra{-} - \ra{+}$ (Supplementary Section 1). This allows us to express the expected collective dynamics of an $N_0$ cell \textit{population} as 

\begin{equation}
\N_A(t)= N_0 \int_{0}^\infty\int_{0}^\infty e^{(y - x) t} P^A_r(x,y;t = 0) \dif x \dif y,
\label{NtHeterogeneous}
\end{equation}
where we sum over all the instantaneous temporal trajectories of all cell types under treatment $A$. We can express the integral in (\ref{NtHeterogeneous}) in a form resembling Eq.\ (\ref{ExponentialDecay}), by writing it as (Supplementary Section 2)  
\begin{equation}
\N_A(t)= N_0 e^{K^A_\lambda(-t)}.
\label{NtCumulant}
\end{equation}

Here $\lambda$ is no longer a \textit{parameter}, but rather it is a \textit{random variable} $\lambda = \ra{-} - \ra{+}$, which is extracted from the probability density $P^A_\lambda(z;t = 0) = \int_0^\infty P^A_r(x,x - z;t = 0) \dif x$ (Fig.\ \ref{Fig1}c). Accordingly, the argument in the exponential expression in (\ref{ExponentialDecay}) is also changed:\ instead of $- \lambda t$, as appears in the classic equation, we now have $\lambda$'s cumulant generating function $K^A_\lambda(t) = \ln [\E(e^{\lambda t})]$ in its place. Hence, treatment $A$, and its $\ra{\pm}$ density $P^A_r(x,y;t = 0)$, can be more naturally characterized by its subsequent $\lambda$ distribution, $P^A_\lambda(z;t = 0)$, which Eq.\ (\ref{NtCumulant}) encapsulates within $K^A_\lambda(t)$. 

Equation (\ref{NtCumulant}), our first key observation, generalizes the classic exponential decay of (\ref{ExponentialDecay}) to treat a population with distributed $\lambda$. It takes the initial population size $N_0$, and its initial $t = 0$ treatment-dependent density function $P^A_r(x,y;t = 0)$ as input, and provides the expected evolution of $\N_A(t)$ as output. It offers a natural generalization in which the linear argument $- \lambda t$, appearing in Eq.\ (\ref{ExponentialDecay}), is substituted by the cumulant generating function $K^A_\lambda(-t)$, which incorporates higher powers of $t$, beyond linear. To observe this, we expand $K^A_\lambda(-t)$ as a power-series in the form
\begin{equation}
K^A_\lambda(-t) = \sum_{n = 1}^\infty \dfrac{1}{n!} \kappa_n (-t)^n = 
- \E(\lambda) t + \dfrac{1}{2} {\rm V}(\lambda) t^2 + \dots,
\label{CumPowerSeries}
\end{equation}

where the coefficients are expressed via $P^A_\lambda(z;t = 0)$'s cumulants $\kappa_n$. To gain intuitive insight, we write the first two term of the series explicitly (r.h.s.), expressing the first cumulant, $\kappa_1$, as the expectation of $P^A_\lambda(z;t = 0)$, and the second cumulant, $\kappa_2$, as its variance, thus grounding $K^A_\lambda(-t)$ in the known moments of the $\lambda$ distribution (note, these are the expectation/variance of $P^A_\lambda(z;t = 0)$, not of $P^A_t(N = n)$, as appear, \textit{e.g}., in Eq.\ (\ref{NtMargins})). Hence, under a homogeneous population in which $\lambda = \E(\lambda)$ and all higher cumulants $\kappa_2,\kappa_3,\dots$ vanish, we retrieve the simple form of (\ref{ExponentialDecay}). However, for all other non-trivial $P^A_\lambda(z;t = 0)$, the higher cumulants, such as the variance ($\kappa_2$), the third central moment ($\kappa_3$) etc., all begin to play a role in shaping $\N_A(t)$, via Eqs. (\ref{NtCumulant}) and (\ref{CumPowerSeries}). This gives rise to nonlinear terms of the form $\sim t^n$ that potentially dominate the dynamics in the limit of large $t$ {\color{blue} $\bullet$} Below we use this formulation to examine the resulting long terms behaviors of $\N_A(t)$, finding that, indeed, these higher cumulants may change the fate of the cell population. 

\begin{figure}[h!]
\centering
\includegraphics[width=0.6\textwidth]{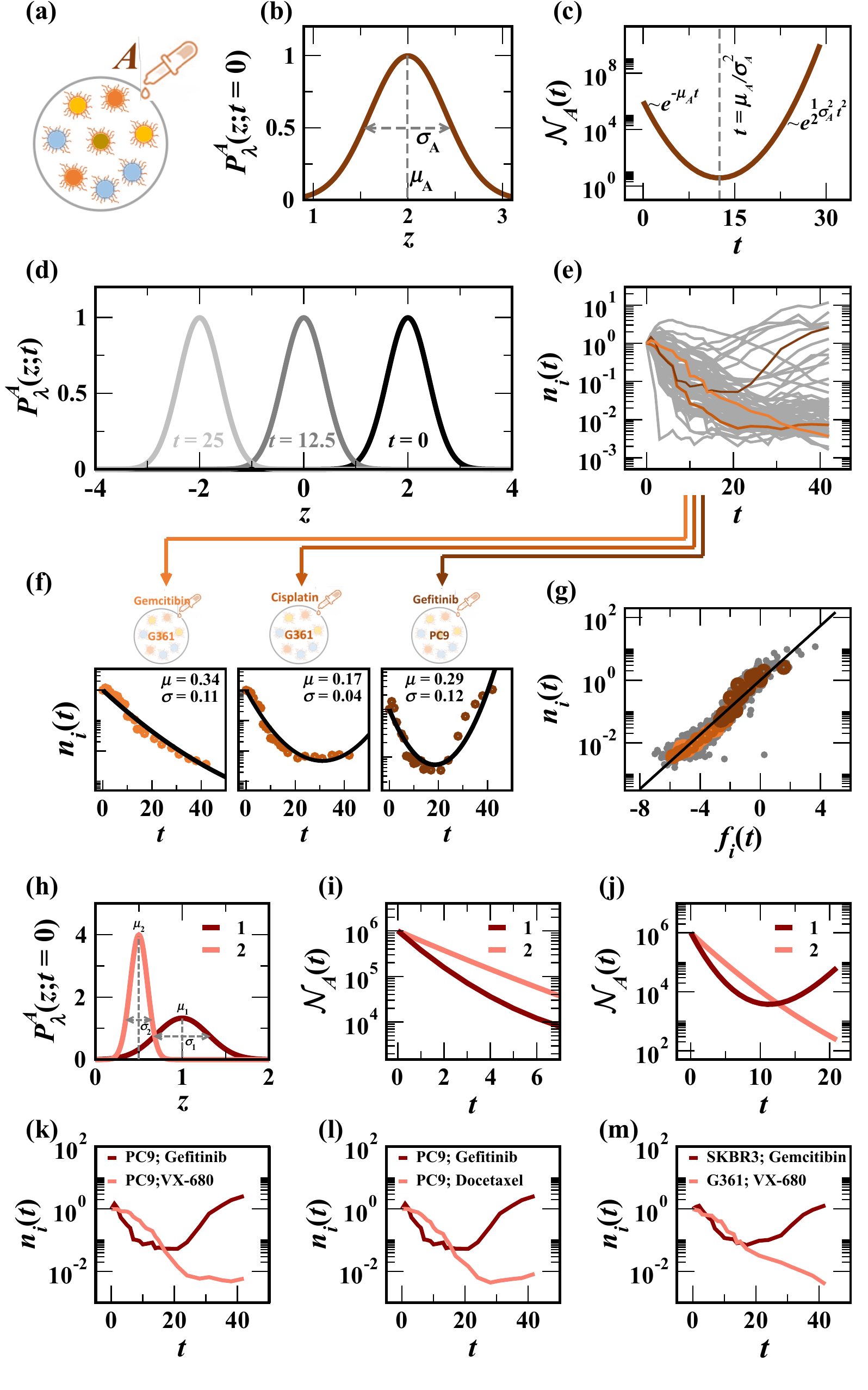}
\vspace{-4mm}
\caption{\footnotesize \textbf{\color{blue} Expected trajectory $\N_A(t)$ under normally distributed rates}.\
(a) - (b) We consider treatment $A$, which invokes a normally distributed response $\lambda \sim \Norm(\mu_A,\sigma_A^2)$. Hence it can potentially include a mixture of positive/negative $\lambda$, as captured by blue/red cells in panel (a).
(c) $\N_A(t)$ vs.\ $t$ begins with an exponential decay driven by the average $E(\lambda) = \mu_A$, but at $t = \mu_A/\sigma_A^2$ begins to reemerge, leading to a long term proliferation driven by $V(\lambda) = \sigma_A^2$. 
(d) The density function $P_\lambda^A(z;t)$ at three different time points $t = 0, 12.5$ and $25$. As the population dynamics progresses the rapidly decaying lineages (large $\lambda$) are eliminated, and the $\lambda$ distribution progressively tends towards the longer living cells (smaller and potentially negative $\lambda$). Hence, at $t \to \infty$ the initially negligible sub-population of $\lambda$-negative cells begins to dominate the population dynamics.
(e) We collected data from $78$ empirical population response patterns ($i = 1,\dots,78$). In experiment $i$ an initial population of $N_{i0}$ cancer cells were subject to a specific chemotherapeutic agent (grey). Here we display the normalized population $n_i(t) = \N_i(t)/N_{i0}$ as obtained from the $i$th cancer/treatment combination. Three specific experimental combinations are highlighted (red shades). For example, in dark red we track cell-line PC9's response to Gefitinib, as represented by the image pointed to by the dark red arrow. The other two combinations are also shown.
(f) We focus on the three highlighted $n_i(t)$ plots of panel (e). Despite the diversity of the observed $n_i(t)$ (circles), we find that they can be well-approximated by Eq.\ (\ref{NtNormal}) under the appropriate selection of $\mu_i,\sigma_i$. 
(g) To test this systematically we plot $n_i(t)$ vs.\ $f_i(t) = -\mu_i t+ \sigma_i^2 t^2/2$. We find that all data-points (grey) of panel (e) collapse onto the universal black solid line predicted by Eq.\ (\ref{NtNormal}). This clearly shows that the diverse response patterns observed across the $78$ experiments all fall within the universal prediction of (\ref{NtNormal}). The three selected $n_i(t)$ from panel (f) are highlighted (red shades).
(h) We consider two competing treatments $A = 1$ (dark red) and $A = 2$ (light red), in which $\mu_1 > \mu_2$ and $\sigma_1 > \sigma_2$. 
(i) The short term response clearly indicates that $1$ is preferable over $2$, as it eliminates the undesired cells more rapidly.
(j) However, in the long term, due to $1$'s higher variance ($\sigma_1 > \sigma_2$), we observe a crossover effect:\ while under $2$ the cells continue to decay, the $1$ treated population changes course, and begins to rapidly proliferate. This highlights the crucial clinical implications of population heterogeneity ($\sigma_A$) over its average ($\mu_A$).
(k) - (m) We observe several instances of such crossover dynamics within our $78$ experimental treatments (see Supplementary Section 6.3).     
}
\label{Fig3}
\end{figure}

\textbf{\color{blue} Long-term behavior of $\N_A(t)$}.\
As our first example we consider normally distributed rates $\lambda \sim \Norm(\mu_A,\sigma_A^2)$ (Fig.\ \ref{Fig3}a,b). Here the first cumulant is $\kappa_1 = \mu_A$ and the second is $\kappa_2 = \sigma_A^2$; the remaining cumulants $\kappa_3,\kappa_4,\dots$ all vanish. Therefore, Eqs.\ (\ref{NtCumulant}) and (\ref{CumPowerSeries}) predict 

\begin{equation}
\N_A(t) = N_0 e^{-\mu_A t+ \frac{1}{2}\sigma_A^2 t^2},
\label{NtNormal}
\end{equation}

an initial decay at a rate $\mu_A$, followed by super-exponential growth, driven by $\sigma_A$ for $t > \mu_A / \sigma_A^2$ (Fig.\ \ref{Fig3}c). Indeed, at the first stages of the population's response, \textit{i.e}.\ small $t$, we observe the \textit{average} decay at a rate $e^{-\mu_A t}$. Yet as time progresses, the rapidly decaying cells are eliminated and the remaining population becomes biased towards the long-living cells (small $\lambda$), hence exhibiting a gradual slowdown in the rate of decay (Fig.\ \ref{Fig3}d). Eventually, at $t > \mu_A / \sigma_A^2$, the surviving population becomes dominated by the cells whose $\lambda < 0$, \textit{i.e}.\ the ones who exhibit net positive growth, and we begin to observe an exponential proliferation. This growth, driven by the \textit{non-average} cells, is, indeed, independent of $\mu_A$, led by the $\lambda$-variance $\sigma_A^2$. It approaches, as $t \to \infty$, $\N_A(t) \sim e^{\sigma_A^2 t^2 / 2}$ - a limit in which the dynamics become fully dominated by the tail of $P^A_\lambda(z;t = 0)$.

To examine prediction (\ref{NtNormal}) in an empirical setting we collected data \cite{Jacob2021} on the response of $3$ different cancerous cell lines under $10$ different chemotherapeutic agents (Supplementary Section 6.3). Testing $6$ different doses of each agent, we arrive at a total of $180$ experimental observations of $N_t$. With each experiment conducted twice, we can average the two observed trajectories to obtain an empirical estimation of $\N_A(t)$, under different treatments $A$. After eliminating all experiments in which the cell population showed no response to the treatment due to low dosage, we remain with $78$ relevant observations (Fig.\ \ref{Fig3}e, Supplementary Section 6.3). At first glance, these $78$ trajectories seem highly diverse and unpredictable. However, Eq.\ (\ref{NtNormal}) suggests that they can all be matched by a single universal function, characterized by two treatment-specific parameters:\ $\mu_A$ and $\sigma_A^2$. To demonstrate this, in Fig.\ \ref{Fig3}f we selected three representative examples, showing that, indeed, by appropriately tuning these two free parameters, their observed trajectories can be well-approximated by Eq.\ (\ref{NtNormal}).

To systematically test prediction (\ref{NtNormal}) across the entire dataset, we extracted the best-fitting $\mu_i,\sigma_i^2$ for all $78$ trajectories ($i = 1,\dots,78$). Namely, we fit the $i$th experiment to the predicted curve $n_i(t) = \N_i(t) / N_{i0} = e^{f_i(t)}$, where $f_i(t) = -\mu_i t+ \sigma_i^2 t^2/2$. In Fig.\ \ref{Fig3}g we collapse all data points by plotting $n_i(t)$ vs.\ $f_i(t)$, which according to our predicted (\ref{NtNormal}) should follow a linear plot (in semi-logarithmic scale). Strikingly, we find that the diversity of Fig.\ \ref{Fig3}e, indeed, condenses quite tightly around our predicted straight line (solid black line), indicating that our theoretically derived $\N_A(t)$ finds clinical relevance in the distributed response of cancer cells to the examined treatments.

\textbf{\color{blue} Crossover effect}.\
The example above emphasizes the crucial role of the $\lambda$-variability ($\sigma_A^2$) in the population dynamics. It also indicates the potential disconnect between the short and long-term behavior of $\N_A(t)$. These two factors together have important clinical implications. To observe this consider two competing treatments, such as two different therapeutics $A = 1$ and $A = 2$, designed to fight a cancerous tumor. Under treatment $1$ the cancer cell population decays via $\lambda_1 \sim \Norm(\mu_1,\sigma^2_1)$ while treatment $2$ invokes a response characterized by $\lambda_2 \sim \Norm(\mu_2,\sigma^2_2)$ (Fig.\ \ref{Fig3}h). 

It is common practice to select the treatment that \textit{kills} the tumor more effectively, \textit{i.e}.\ leads to a more rapid decay of $N_t$. To test this, one subjects a sample of $N_0$ cells to both $1$ and $2$ and observes the exponential decline in $N_t$ under the two  conditions. Here, for example, $\mu_1 > \mu_2$ and hence, throughout the first few days it seems quite clear that $1$ represents the preferable treatment (Fig.\ \ref{Fig3}i). The challenge is that, as Eq.\ (\ref{NtNormal}) indicates, the short-term behavior only expresses the population's \textit{average} decay rate $\mu_1,\mu_2$, but offers no information on its \textit{variance}. In contrast, the long-term behavior of $N_t$ is primarily driven precisely by $\sigma_1$ and $\sigma_2$, and therefore, selecting $1$ over $2$ based on the observed $\mu_1 > \mu_2$, \textit{i.e}.\ the \textit{average} $\lambda$s, may potentially overlook the impact of the $\lambda$-\textit{variability}.   

Indeed, here, since $\sigma_1^2 > \sigma_2^2$, we observe a crossover (Fig.\ \ref{Fig3}j):\ while $1$ seems preferable over $2$ in the short run, after some time the roles are reversed, the $1$ treated population begins to proliferate, while the cells subject to $2$ continue to decay {\color{blue} $\bullet$} This has crucial clinical consequences, highlighting that when designing treatment, such as chemotherapy, one must consider the full $\lambda$-distribution, or at the least its variance, and not just the population average.

To observe this in our clinical setting, we return to our $78$ empirical population dynamics. As predicted, we found among these experiments several instances of the crossover effect, where one treatment seemed preferable within the first few days (Fig.\ \ref{Fig3}k-m, dark red), only to be overcome by its initially slower contender after $\sim 2$ weeks (light red). This demonstrates the deep clinical implications of our predicted Eq.\ (\ref{NtCumulant}), showing that $\lambda$-heterogeneity plays a crucial role in the long-term population response. We, therefore, recommend to rank therapeutics not just based on the average cellular responses that they invoke, but also based on their response variability - here the distributed birth/death rates.

\textbf{\color{blue} Averting recurrence - the role of deviations from $\N_A(t)$}.\
A deeper look into Eq.\ (\ref{NtNormal}) and its consequences in Fig.\ \ref{Fig3} seems to suggest that, regardless of their specific parameters, all treatments are deemed to fail, unless they have strictly $\sigma_A = 0$, \textit{i.e}.\ a completely homogeneous $\lambda$ across all cells. Indeed, as long as there exists a fraction of the population with positive growth ($\lambda < 0$), the eventual recurrence at $t \to \infty$ is inevitable. How then will any treatment enable long term remission?

This seemingly inevitable recurrence is a consequence of two simplifications employed in the above analysis:\ first, taking $P^A_\lambda(z;t = 0)$ to follow a perfect normal distribution, under which $\lambda \in (-\infty,\infty)$, second, focusing on the \textit{expected trajectory} $\N_A(t)$, while disregarding the stochastic nature of $N_t$, whose behavior may deviate from $\N_A(t)$ as discussed in Eq.\ (\ref{NtMargins}). Hence, as our next step, we turn to observe richer patterns in $\N_A(t)$, including ones without recurrence, enabled by generalizing $P^A_\lambda(z;t = 0)$ beyond the normal distribution. We then advance further to calculate $\V_A(t)$, helping us capture stochastic deviations from the expected trajectory, which - we find - have a crucial impact on the population's long-term prevalence, and hence on the potential success and efficiency of treatment.

\vspace{5mm}
\textbf{\color{blue} \large Classes of population dynamics}

As our example above indicates, the cell-population $\N_A(t)$ tends, with time, to become dominated by the persistent cell lineages, namely the ones with the smallest $\lambda$. In a finite population $N_0$, the longest lasting lineage is, typically, the one characterized by $\lmin$. We therefore, consider a family of $\lambda$-density functions that can be expanded around this minimal $\lambda$. Quite generally, this can be expressed as a power series of the form \cite{Hahn1995}
\vspace{1mm}
\begin{equation}
P^A_\lambda(z; t = 0) = \sum_{n = 0}^{\infty} 
C_n (z - \lmin)^{\Psi_n},
\label{Hahn}
\vspace{-2mm}
\end{equation}

capturing a broad family of potential density functions. In (\ref{Hahn}), the powers $\Psi_n \in \mathbb{R}$ represent a series of real powers arranged in ascending order, \textit{i.e}.\ $\Psi_{n + 1} > \Psi_n$. Hence, it captures a generalization of the classic Taylor expansion, in which, in addition to natural powers, we allow also for negative and non-integer $\Psi_n$. The existence of $\lmin$ is guaranteed as long as $N_0$ is finite. Note that, in principle, the parameters $C_n,\Psi_n$ and $\lmin$ are $A$-dependent, \textit{i.e}.\ $C_{A,n},\Psi_{A,n},\lambda_{A,{\rm min}}$. However, to refrain from cumbersome notations, we omitted the identifier $A$, keeping this dependence implicit.

To predict $\N_A(t)$ via (\ref{NtCumulant}) we show in Supplementary Section 3 that (\ref{Hahn})'s cumulant generating function follows
\begin{equation}
K^A_\lambda(-t) = - \lmin t + 
\ln \left(\sum_{n = 0}^\infty C_n \Gamma(\Psi_n + 1) t^{-\Psi_n - 1} \right)
\label{HahnCumulant}
\end{equation} 

where $\Gamma(x)$ represents the continuous extension to the factorial operation, \textit{i.e}.\ $\Gamma(x) = \int_0^\infty \xi^{x - 1} e^{-\xi} \dif \xi$. In the limit $t \to \infty$ we only preserve the leading powers on the r.h.s., providing us with $\lim_{t \to \infty} K^A_\lambda(-t) = - \lmin t + \ln [C_0 \Gamma(\Psi_0 + 1) t^{-\Psi_0 - 1}]$. Substituting this into (\ref{NtCumulant}) leads to

\begin{equation}
\N_A(t \rightarrow \infty) = 
N_0 C_0 \Gamma(\alpha) t^{-\alpha} e^{-\lmin t},
\label{NtInfty}
\end{equation}

where $\alpha = \Psi_0 + 1$. Therefore, starting from an initial population of $N_0$ cells, whose $\lambda$-distribution is taken from (\ref{Hahn}), we arrive at $t \to \infty$ at a population driven by $\lmin$ and by the leading power $\Psi_0$. These two parameters of (\ref{HahnCumulant}) cover the limit of the slowest decaying cell lineages from $N_0$, namely those with $\lambda = \lmin$ and their immediate neighbors, captured by $(z - \lmin)^{\Psi_0}$. These slow lineages at the $z \to \lmin$ boundary of $P^A_\lambda(z;t = 0)$ are, indeed, the last to prevail.        

Equation (\ref{NtInfty}) represents our second key result, describing $\N(t)$'s asymptotic behavior under the general response distribution of Eq.\ (\ref{Hahn}). Depending on $\lmin$ it predicts three distinct classes of population dynamics, whose clinical implications we investigate below:\

\textit{\color{blue} $\bullet$ Exponential remission} ($\lmin > 0$, Fig.\ \ref{Fig2}a-c).\  
In case $\lmin > 0$, then at large $t$ all rapidly decaying cells have already been eliminated, and the remaining sub-population comprises only the long-living lineage with the smallest $\lambda$. This is captured in Eq.\ (\ref{NtInfty}) by the exponential term $e^{-\lmin t}$, which dominates over the polynomial $t^{-\alpha}$. We observe this form of dynamics in the classic case of bacterial persistence, in which $P^A_\lambda(z;t = 0)$ is split into two sub-populations, a rapidly decaying majority and a persistent minority. \cite{Balaban2004,Keren2004} At large $t$, as we are only left with the persistent cells we, indeed, observe the slow decaying branch of $\N_A(t) \sim e^{-\lmin t}$. Still, despite being the slower branch, it continues to exhibit an exponential decay, hence ensuring a bounded remission timescale of the form (\ref{TLogN0}).

\textit{\color{blue} $\bullet$ Slow remission} ($\lmin = 0$, Fig.\ \ref{Fig2}d-f).\ 
This class covers the many relevant density functions, from exponential to Boltzmann, that are characterized by strictly positive $\lambda$, \textit{i.e}.\ $P^A_{\lambda}(z < 0;t = 0) = 0$. These systems are, therefore, bounded by $\lmin = 0$. In clinical terms, this represents conditions under which cells can, at most, overcome the adverse effects of the treatment, yet their reproduction is still suppressed, and hence they exhibit no net growth. Under such conditions, Eq.\ (\ref{NtInfty}) predicts 

\vspace{-2mm}
\begin{equation}
\N_A(t \to \infty) \sim t^{-\alpha},
\label{LongTail}
\end{equation}

a polynomial tail, which captures an extremely slow population decay. Such populations, after an initial short-lived, exponential phase, will enter a power-law dynamics, and hence will lack a typical extinction timescale, thus diverting from Eq.\ (\ref{TLogN0})
 
\textit{\color{blue} $\bullet$ Recurrent} ($\lmin < 0$, Fig.\ \ref{Fig2}g-i).\
The third option is that $\lmin$ is negative, and hence, even under the treatment $A$, there continues to be a finite probability for positive growth. The expectation value $\N_A(t)$, therefore, exhibits an exponential proliferation for large $t$, as the initially marginal $\lambda$-negative fraction of cells, eventually gains dominance over the population. In our analysis below, we show that while this captures the \textit{expectation} value $\N_A(t)$, in practice, there exist broad conditions under which the recurrence probability is small. Namely that a significant fraction of the \textit{stochastic} realizations avoids this reemergence. Hence, the recurrent class captures a \textit{possibility} for population recovery, but not necessarily an inevitability of such fate.  

Therefore, to predict the actual observed outcome under the different classes, we advance beyond the analysis of $\N_A(t)$ and introduce stochasticity via the variance $\V_A(t)$. Since the exponential class was already broadly studied in the context of bacterial persistence, \cite{Moyed1983,Korch2003,Kussell2005,D2009,Hammoud2013} we direct our focus to the remaining classes:\ slow and recurrent, whose dynamics are fundamentally distinct from the classic Eqs.\ (\ref{ExponentialDecay}) and (\ref{TLogN0}). In both classes, we seek the probability and typical timescales for remission, helping us characterize potentially successful, failed or inefficient treatments.

\vspace{5mm}
\textbf{\color{blue} \large Population extinction - the role of $\V_A(t)$}

Our predictions for $\N_A(t)$ in the different classes, are all rooted in the cumulant-based equation (\ref{NtCumulant}), whose exponential form can never reach the extinction state $\N_A(t) = 0$. This, however, is because $\N_A(t)$ represents the \textit{expectation} of $N_t$, and hence, overlooks the discrete and potentially stochastic nature of the actual observed population dynamics. Indeed, $\N_A(t)$, as defined in our formulation, captures an average over $M \to \infty$ realizations, and can, therefore, assume continuous values, and reach arbitrarily close to zero, without ever going extinct. This is in contrast to the specific stochastic realizations, where extinction is possible, and in some cases even represents the most probable outcome. 

As an example, let us revisit Eq.\ (\ref{NtNormal}) and its two recurrent trajectories $\N_A(t)$ for $A = 1,2$ of Fig.\ \ref{Fig3}h-j. There, while both $1$ and $2$ eventually enter the proliferation phase, the turnover in $2$ occurs much later, at a point where the original population is reduced by a factor of $\sim 10^{-5}$. This is as opposed to $1$, whose turning point occurs at $\sim 5 \times 10^{-3} N_0$. Consequently, if we begin with a finite population of, say $N_0 \sim 10^4$ or $10^5$ cells, treatment $2$ would reach its minimum point when $N_t$ is of order unity, a point from which it may likely transition to $N_t = 0$, thus avoiding the recurrent branch. Under similar conditions treatment $1$'s tipping point occurs far from $N_t = 0$, and hence, while $1$ seems more effective at first, it is more likely than $2$ to result in eventual recurrence. 

The crucial point is that such distinction between treatments $1$ and $2$, whose clinical outcomes are, clearly, consequential, cannot be observed through the \textit{expected trajectories} $\N_A(t)$, which, by design, overlook the stochastic and discrete nature of the random $N_t$. In fact, even an extremely small fraction, say $\sim 1\%$, of realizations of treatment $1$ that cross into the proliferation branch, can push its \textit{expected} trajectory $\N_A(t)$ to diverge. Yet, still, from a clinical perspective, such outcome represents a $99\%$ success rate, which is highly desirable. Hence, $\N_A(t)$'s recurrence does not, on its own, render the treatment unsuccessful. To evaluate $A$'s success, systematically, we must capture the potential deviations from $\N_A(t)$ - precisely the role of the variance $\V_A(t)$.     

The variance, we show in Supplementary Section 2, depends not just on the random variable $\lambda = \ra{-} - \ra{+}$, but also on $\varphi = \ra{-} + \ra{+}$. Therefore it incorporates the joint density function $P^A_{\lambda,\varphi}(z,u;t = 0)$ and its ensuing bi-variate cumulant generating function $K_{\lambda,\varphi}(t,s) = \ln [\E(e^{\lambda t + \varphi s})]$. \cite{Pistone1999} Our analysis helps predict the variance, which we find, follows
\vspace{2mm}
\begin{equation}
\V_A(t) = N_0 \int \left. \left[
2 \dod{\cumtwo{\lambda}{\varphi}{-2t}{s}}{s}
e^{\cumtwo{\lambda}{\varphi}{-2t}{s}}
- \dod{\cumtwo{\lambda}{\varphi}{-t}{s}}{s} 
e^{\cumtwo{\lambda}{\varphi}{-t}{s}}
\right] \right|_{s = 0}
\dif t,
\label{Variance}
\end{equation}

a closed form solution based on $K_{\lambda,\varphi}(t,s)$. Equation (\ref{Variance}) takes a heterogeneous cell population, with any arbitrary $P^A_r(x,y;t = 0)$, and hence $P^A_{\lambda,\varphi}(z,u;t = 0)$, as input, and provides the variance at time $t$ as output. This variance quantifies the level of uncertainty around $\N_A(t)$ as time progresses. Therefore, in each stochastic realization of our heterogeneous population dynamics, we take $\N_A(t)$ of Eq.\ (\ref{NtCumulant}) to capture the expected population at time $t$, and $\V_A(t)$ in (\ref{Variance}) to describe the potential deviations from this expectation (Fig.\ \ref{Fig1}b).

To understand how $\N_A(t)$ and $\V_A(t)$ can predict instances of extinction, consider the \textit{observed} population $N_t$ at any given time. Using Eq.\ (\ref{NtMargins}), we can assume, with relatively high probability, that it is within the interval $\N_A(t)[1 \pm 1/\Q_A(t)]$, where 
\vspace{2mm}
\begin{equation}
\Q_A(t) = \dfrac{\N_A(t)}{\sqrt{\V_A(t)}}.
\label{Qt}
\end{equation}

As long as $\Q_A(t) \gg 1$, $N_t$ is guaranteed, with high probability, to remain at a safe distance from $N_t = 0$. If, however, at a certain instance we have $\Q_A(t) \le 1$ we observe a significant probability of $N_t$ hitting zero, as it fluctuates around the analytically predicted $\N_A(t)$; Fig.\ \ref{Fig1}d.

For example, consider the recurrent population of Fig.\ \ref{Fig2}i. If, \textit{e.g}., at the minimum point, $\N_A(t)$ is of the same order as $\sqrt{\V_A(t)}$, there is a likelihood that a significant fraction of the realizations will hit the absorbing $N_t = 0$. This is precisely captured by $\Q_A(t) \le 1$. As $\Q_A(t)$ decreases, this likelihood grows, until under $\Q_A(t) \ll 1$, the population is almost guaranteed to reach extinction. We emphasize, that, even then, $\N_A(t)$ may continue to diverge, as predicted, for example, in (\ref{NtNormal}), but this divergence will be driven by the increasingly \textit{rare} instances of recurrence, not by the \textit{majority} of extinction events. 

Hence, we arrive at our final key prediction, extracting the statistics of $P^A_{\lambda,\varphi}(z,u;t = 0)$ that help characterize treatment $A$'s efficacy:\

\textit{\color{blue} $\bullet$ Successful treatment}.\
Treatment $A$'s success is predicted by the condition
\vspace{2mm}
\begin{equation}
\begin{array}{ccc}
Q_A \le 1
& \,\,\,\,\,\, &
Q_A \equiv \displaystyle \min_t \Big\{ \Q_A(t) \Big\}
\end{array},
\label{Qmin}
\end{equation}  
namely, that at some point in time $t$, $\Q_A(t)$ runs below unity, and hence, with high probability $N_t$ will hit zero, \textit{i.e}.\ remission. The smaller is $Q_A$, the more probable is the treatment's successful outcome.

\textit{\color{blue} $\bullet$ Efficient treatment}.\
In case (\ref{Qmin}) is satisfied and the treatment is predicted successful, the question that remains is what will be the expected treatment duration. This is captured by 

\begin{equation}
T_A = \min_t \Big \{ t \Big | \Q_A(t) \le 1 \Big \},
\label{T}
\end{equation}   

namely the earliest point in time by which the system satisfies condition (\ref{Qmin}). The smaller is $T_A$, the more efficient is treatment $A$. 

\begin{Frame}
We have now arrived at our complete predictive framework for heterogeneous population dynamics. We begin with a population of $N_0$ adversarial cells (bacteria, tumor etc.), which we subject to therapeutic $A$. The population's response to $A$ is characterized by $P^A_{\lambda,\varphi}(z,u;t = 0)$, a bi-variate density function that helps characterize the (stochastic) birth/death rates of all cells under $A$ 
{\color{blue} $\bullet$} From these initial conditions we predict the expected population $\N_A(t)$ via (\ref{NtCumulant}) and its variance $\V_A(t)$ via (\ref{Variance}). The former captures the average outcome of treatment $A$; the latter, its potential variability across different trials 
{\color{blue} $\bullet$} We can now use both predictions to evaluate $\Q_A(t)$ in (\ref{Qt}) and asses treatment $A$'s potential merit:\ Its \textit{success} likelihood is given by condition (\ref{Qmin}) and its \textit{efficiency} by (\ref{T}) 
{\color{blue} $\bullet$} Our analysis indicates two potential dynamic classes that we wish to avoid:\ (i) \textit{Recurrent}, in which the population springs to back to exponential growth, representing an \textit{unsuccessful} outcome; (ii) \textit{Slow remission}, in which the time to reach extinction is potentially divergent, capturing an \textit{inefficient} outcome 
{\color{blue} $\bullet$} Taken together, our predictive framework accepts $N_0$, the initial cell-population, and $P^A_{\lambda,\varphi}(z,u;t = 0)$, which characterizes treatment $A$'s clinical profile, and ensures, \textit{a priori}, that we select a treatment that minimizes both of these risks. 
\end{Frame}

\vspace{2mm}
\textbf{\color{blue} \large Treatment success and efficiency}
 
To test our predictions we constructed an array of $1.1 \times 10^4$ \textit{in silico} experiments. We begin with cellular populations whose sizes range from $N_0 = 10^3$ to $N_0 = 10^7$, and examine their dynamics under different therapeutics, each characterized by its distinct $P^A_{\lambda,\varphi}(z,u; t = 0)$. For each experiment we conduct $M = 100$ stochastic trials. This allows us to evaluate $\N_A(t)$, the expectation, through the observed average population dynamics $\av{N_t}$, which we extract from the actual stochastic runs. Similarly we evaluate $\V_A(t)$, the variance, through $\av{N_t^2} - \av{N_t}^2$, again - calculated from the repeated stochastic simulations. With this at hand we examine conditions (\ref{Qmin}) and (\ref{T}) to test their predictive power for the treatment success/efficiency.   

\vspace{2mm}
\textbf{\color{blue} Slow remission}.\
To examine the slow class we consider the Gamma density function
\vspace{2mm}
\begin{equation}
P^A_\lambda(z; t = 0) = C z^{\alpha - 1}e^{-l z} \,\,\,\,\,\,\,\,\,\,\, z \ge 0; \,\,\,\, \alpha,l > 0,
\label{Gamma}
\end{equation}

representing a broad family of strictly positive $\lambda$ distributions, which is frequently encountered in relevant contexts. \cite{Atia2018,Tkachenko2020,Hogg1978} For example, setting $\alpha = 1$ we have $P^A_\lambda(z;t = 0) \sim {\rm Exp}(l)$, an exponential distribution; alternatively under $l = 1/2$ and $\alpha = k/2$ we arrive at the classic $\chi^2$-distribution; finally, for $\alpha = 3/2$ we obtain the Boltzmann distribution. In (\ref{Gamma}) the normalization coefficient is $C = l^\alpha/ \Gamma(\alpha)$, the expectation is $\E(\lambda) = \alpha/l$ and the variance is ${\rm V}(\lambda) = \alpha/l^2$. For simplicity, in (\ref{Gamma}) we only show the marginal $\lambda$ distribution, leaving the analysis of the complete $P^A_{\lambda,\varphi}(z,u;t = 0)$ for the supplement (Supplementary Section 5). 

In Supplementary Section 5 we use our formalism to show that under $P^A_\lambda(z;t = 0)$ of the form of (\ref{Gamma}) the population dynamics asymptotically follow 
\vspace{2mm}
\begin{equation}
\begin{array}{ccccc}
\N_A(t) \sim N_0 t^{-\alpha};
& \,\,\,\,\, &
\V_A(t) \sim N_0 t^{1 - \alpha}; 
& \,\,\,\,\, &
\Q_A(t) \sim \sqrt{N_0 t^{- \alpha - 1}}
\end{array}.
\label{NtGamma}
\end{equation}  

This captures precisely the predicted \textit{slow}, sub-exponential, dynamics, in which the population converges to a power-law tail, in accordance with prediction (\ref{LongTail}). The exponent, characterizing this long-term behavior, depends solely on the parameter $\alpha$ in (\ref{Gamma}). Therefore, regardless of the specific assignment of $\ra{+}$ and $\ra{-}$ via $P^A_r(x,y;t = 0)$, or of the resulting complete bi-variate density function $P^A_{\lambda,\varphi}(z,u;t = 0)$, the population decay, in this class of systems, is asymptotically driven only by $P^A_\lambda(z;t = 0)$, and specifically only by its parameter $\alpha$.   

\begin{figure}[h!]
\centering
\includegraphics[width=0.7\textwidth]{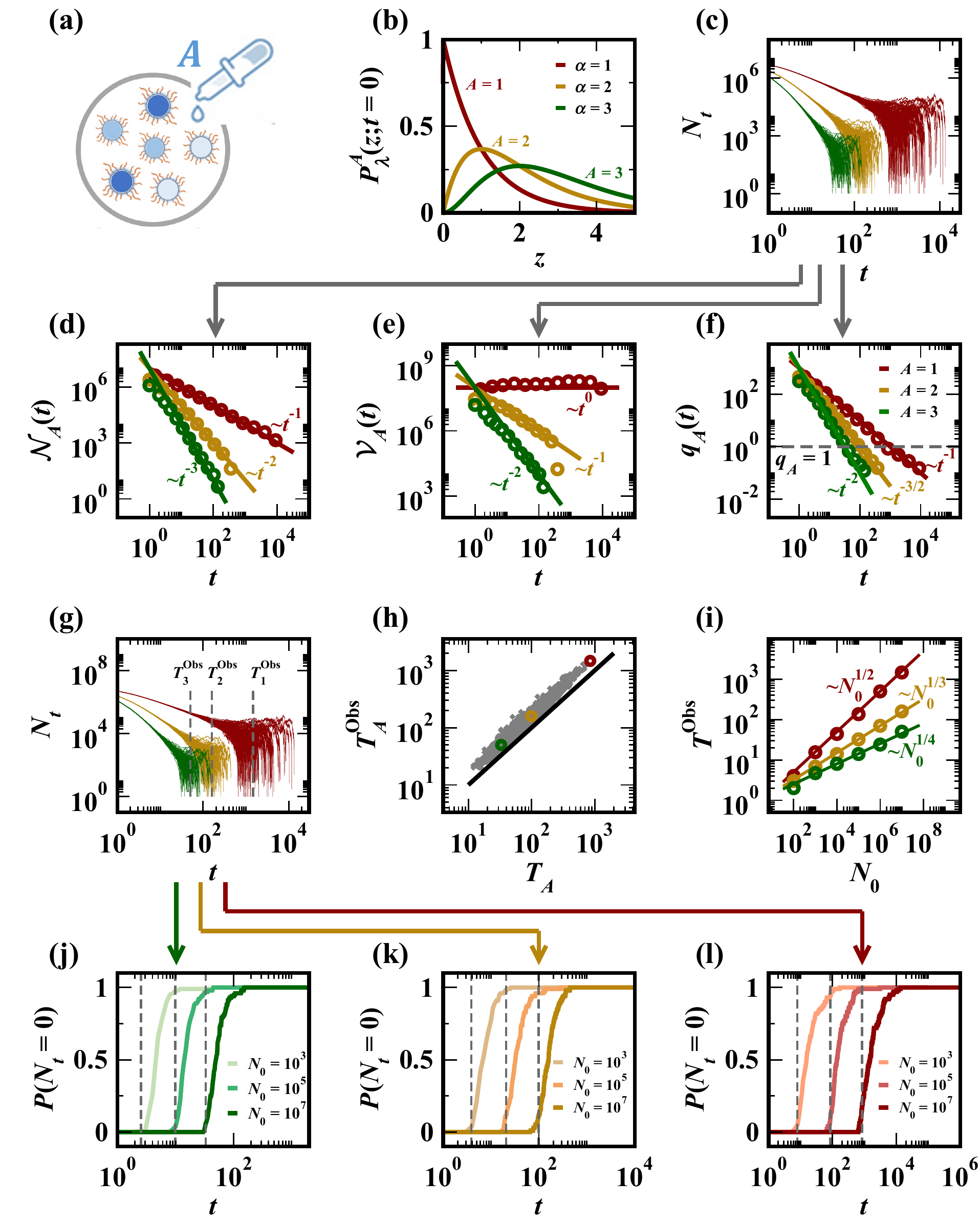}
\vspace{-5mm}
\caption{\footnotesize \textbf{\color{blue} Slow remission}.\
(a) - (b) To observe our slow remission class we consider three competing treatments $A = 1,2,3$, each characterized by a different $\Gamma(\alpha,l)$-distributed response pattern ($\alpha = 1,l = 1$, red; $\alpha = 2,l = 1$, orange; $\alpha = 3,l = 1$, green). 
(c) The observed population $N_t$ vs.\ $t$ as obtained under treatments $1$ (red), $2$ (orange) and $3$ (green). For each treatment we show $M = 100$ independent realizations, starting from $N_0 = 10^7$. 
(d) Averaging over all $M$ realizations we extract $\N_A(t)$. As predicted in (\ref{NtGamma}), we find that $\N_A(t) \sim t^{-\alpha}$ for $\alpha = 1,2,3$ (red, orange and green solid lines, respectively). This confirms our predicted slow power-law decay.
(e) $\V_A(t)$ vs.\ $t$ (circles) as obtained from the realizations $N_t$ of panel (c). Once again, we observe the scaling (solid lines) predicted in Eq.\ (\ref{NtGamma}).
(f) $\Q_A(t)$ vs.\ $t$ (circles), alongside our predicted scaling (solid lines). Our analysis predicts that the time $T_A$ for the treatment to reach success occurs when $Q_A \le 1$ (horizontal dashed line). Hence the intersection between $\Q_A(t)$ (colored lines) and the horizontal dashed line represents the \textit{expected} population extinction time $T_A$.
(g) Using our stochastic realizations of panel (c) we extract the \textit{observed} extinction time $T_A^{\rm Obs}$ ($A = 1,2,3$) as the time when half of all realizations reached $N_t = 0$, namely when $P(N_t = 0) = 1/2$ (vertical dashed lines). Indeed, for $t > T_A^{\rm Obs}$ extinction becomes the more likely outcome, hence capturing the practically observed typical treatment duration. In our logarithmic axes, since $N_t = 0$ cannot be displayed, extinction is depicted by the sharp drops in $N_t$ observed towards the r.h.s.\ of each plot. 
(h) The observed extinction time $T_A^{\rm Obs}$ vs.\ the predicted $T_A$ in Eq.\ (\ref{T}) as evaluated using the dashed line intersection of panel (f). We observe a striking agreement between observation and theory, sustained over a set of $4,400$ distinct combinations of initial conditions ($N_0 \in (10^3,10^7)$) and treatment profiles ($\alpha \in (1,4)$); grey dots. The three specific examples of panels (c) - (g) are highlighted via the red, orange and greed dots. This quite clearly indicates the predictive power of $T_A$ in (\ref{T}) to extract the actual treatment duration under a diverse set of fully stochastic settings. 
(i) $T_A^{\rm Obs}$ vs.\ the initial population size $N_0$ under treatments $A = 1,2,3$ (circles). Our theoretical predictions of Eq.\ (\ref{TGamma}) are also shown (solid lines). Indeed, we find that the treatment duration scales with $N_0$, precisely as predicted. This directly demonstrates the \textit{slow remission}, which diverges polynomially with the malignant population size.
(j) The success probability $P(N_t = 0)$ vs.\ $t$ under treatment $3$ with initial population sizes $N_0 = 10^3,10^5$ and $10^7$. We find that, indeed, the success probability jumps abruptly from zero to unity around the predicted $T_A$ under each scenario (vertical dashed lines). This indicates that, indeed, treatment duration is highly concentrated around the statistic $T_A$ in Eq.\ (\ref{T}). 
(k) - (l) Similar results are also observed under treatments $2$ (orange) and $3$ (red). 
}
\label{Fig4}
\end{figure}

To examine the predictions in (\ref{NtGamma}) we numerically simulated $4,400$ different treatment scenarios. In each simulation we begin with $N_0 \in (10^3,10^7)$ cells and assign their birth/death rates $\ra{+}$ and $\ra{-}$ at random from $P^A_r(x,y;t = 0)$; see Supplementary Section 6.2 where we link the $\ra{+},\ra{-}$ distribution to (\ref{Gamma}). We focus on three separate $\lambda$-distributions in the form (\ref{Gamma}), as shown in Fig.\ \ref{Fig4}a,b:\ Treatment $1$ having $\alpha = 1$ (red); $2$ with $\alpha = 2$ (orange); and $3$ for which we set $\alpha = 3$ (green). In each simulation we track the random variable $N_t$ across $M = 100$ stochastic realizations (Fig.\ \ref{Fig4}c), and measure its average to evaluate $\N_A(t)$ and variance to extract $\V_A(t)$ (Fig.\ \ref{Fig4}d,e, circles). These observations, we emphasize, capture the complete stochastic population dynamics, as described in Fig.\ \ref{Fig1}. Namely, the original population rates are selected at random, and their temporal evolution is modeled via a series of stochastic replication and elimination events.

As predicted in (\ref{NtGamma}) we find that our simulation results condense along three separate branches, each corresponding to a different $\alpha$ value. The average population $\N_A(t)$ decays as $\sim t^{-\alpha}$ and the variance $\V_A(t)$ as $\sim t^{1 - \alpha}$ (solid lines), in perfect agreement with our analytical predictions in Eq.\ (\ref{NtGamma}). In Fig.\ \ref{Fig4}f we further measure $\Q_A(t)$ using (\ref{Qt}), finding that it too follows after our prediction in (\ref{NtGamma}). These findings clearly show that our analytical framework, and specifically Eqs.\ (\ref{NtCumulant}) and (\ref{Variance}), can indeed capture the actual population response patterns. Therefore, while the observed response trajectories $N_t$ are random and fluctuative, their stochastic nature can be well-approximated by our analytically predicted $\N_A(t),\V_A(t)$ and $\Q_A(t)$.      

\textbf{\color{blue} Efficiency}.\
Under the asymptotic behavior predicted in (\ref{NtGamma}) any treatment within the form (\ref{Gamma}) is guaranteed to be successful, as, indeed, we have $\N_A(t \to \infty) \to 0$. Hence, the only risk is that, while successful, $A$ may be inefficient, taking an extremely long time to eliminate the cell population. To quantify this potential inefficiency, we use condition (\ref{T}) to asses the typical time for population extinction (Fig.\ \ref{Fig4}f, dashed line). Of course, given the random nature of $N_t$, we expect the actual extinction time to vary across realizations, and therefore we define the observed extinction time $T_A^{\rm Obs}$, as the time when $P(N_t = 0) \ge 1/2$, namely the time-point beyond which extinction becomes more likely than survival. We can extract $T_A^{\rm Obs}$ from our simulation results by seeking the first point in time in which $50\%$ of the $N_t$-realizations reached $N_t = 0$, \textit{i.e}.\ the observed median extinction time (Fig.\ \ref{Fig4}g, vertical dashed lines). We then compare this observed elimination time to our predicted $T_A$ in Eq.\ (\ref{T}). Plotting $T_A^{\rm Obs}$ vs.\ $T_A$ across all $4,400$ scenarios, we find a striking agreement, demonstrating the predictive power of our condition (\ref{T}) to capture the actual timescales for treatment success (Fig.\ \ref{Fig4}h). 

To gain deeper insight we focus on a set of nine specific realizations, representing a range of population sizes ($N_0 = 10^3,10^5,10^7$) combined with the different treatment profiles ($A = 1,2,3$). In each case we present $P(N_t = 0)$ vs.\ $t$, capturing the probability of treatment success vs.\ time (Fig.\ \ref{Fig4}j-l). Indeed, we find that the success probability jumps from zero to unity quite rapidly, precisely as we cross the $\Q_A(t) = 1$ mark (vertical dashed lines), \textit{i.e}.\ condition (\ref{T}).    

These observations establish that (i) $\Q_A(t)$ follows the analytical prediction of Eq.\ (\ref{NtGamma}), as shown in Fig.\ \ref{Fig4}f; (ii) that it can, indeed, be used as a systematic predictor of the treatment duration through the condition $\Q_A(t) \le 1$ (Fig.\ \ref{Fig4}h). We can now combine (i) and (ii) to evaluate the efficiency of treatments against cell populations that fall within the Gamma family of Eq.\ (\ref{Gamma}). Taking $\Q_A(t)$ from (\ref{NtGamma}) and setting it to unity, we extract the predicted extinction time as

\begin{equation}
T_A \sim N_0^{\beta}
\label{TGamma}
\end{equation}   

where $\beta = 1/(1 + \alpha)$, capturing the typical treatment duration required to eliminate a cell-population of size $N_0$. Equation (\ref{TGamma}) is profoundly distinct from the classic logarithmic dependence of Eq.\ (\ref{TLogN0}), portraying an extremely inefficient treatment, whose termination time $T_A$ scales with the initial population size $N_0$. For a sufficiently large $N_0$ the treatment duration in (\ref{TGamma}) diverges polynomially, and hence cannot eliminate the undesired cell population within a bounded timescale.

This divergence with $N_0$ can be directly observed in our nine examples of Fig.\ \ref{Fig4}j-l, where both the theoretical $T_A$ (vertical dashed lines) and the observed $T_A^{\rm Obs}$ (sharp inclines in $P(N_t = 0)$) increase significantly with population size, ranging, \textit{e.g}., from $\sim 2 \times 10^1$ to $\sim 10^3$ as $N_0$ is increased from $10^3$ to $10^7$. To observe this more systematically, we measured the average \textit{observed} treatment duration $T_A^{\rm Obs}$ as we vary $N_0$ across six orders of magnitude (Fig.\ \ref{Fig4}i, symbols). We find that for large $N_0$, the extinction time follows precisely the scaling prediction of (\ref{TGamma}), accurately recovering its $\alpha$-dependent scaling exponent (solid lines).

The distribution in (\ref{Gamma}) represents a broad family of potential density functions that belong to our slow remission class. It was selected here for its anaytical tractability, but can be generalized further by considering all density functions of the form (\ref{Hahn}), having $\lmin = 0$. If a specific treatment $A$ falls within this class, it is guaranteed to be successful, \textit{i.e}.\ lead to $N_t = 0$, but at the same time risks extreme levels of inefficiency, incorporating timescales that diverge with the population size $N_0$. This inefficiency, analytically predicted by our formulation, can only be observed through our probabilistic description, in which $A$ is characterized by $P^A_{\lambda,\varphi}(z,u;t = 0)$. Indeed, had we only considered the population average $\E(\lambda)$, our predictions would align with the rapid exponential decay of (\ref{ExponentialDecay}) and (\ref{TLogN0}). Similarly, using the expectation $\N_A(t)$, and ignoring stochasticity, as captured by $\Q_A(t)$, one cannot derive the actual remission time $T_A$ as appears in (\ref{TGamma}) - a clinically crucial prediction that is unequivocally validated in Fig.\ \ref{Fig4}i. 

Yet, as our results clearly indicate, it is the heterogeneity in the population's response, as captured by (\ref{Gamma}), and its stochastic birth/death dynamics as tracked via $\N_A(t),\V_A(t)$ and $\Q_A(t)$, that together drive the long-term efficacy of $A$. The resulting clinical implications, summarized in Eq.\ (\ref{TGamma}), become especially relevant in the limit of $N_0 \gg 1$, \textit{e.g}., a large tumor, where our analysis indicates that $T_A$ diverges as $N_0^\beta$.         

\textbf{\color{blue} Recurrent dynamics}.\
We now turn to the recurrent class ($\lmin < 0$), revisiting our normally distributed birth/death rates from Fig.\ \ref{Fig3}. Hence, we take $P^A_{\lambda,\varphi}(z,u;t = 0)$ to be a bi-variate normal distribution\cite{Hogg1978}, namely $\lambda,\varphi \sim \Norm(\mu_A,\sigma^2_A;\nu_A,\delta^2_A)$. Our previous analysis, expressed in Eq.\ (\ref{NtNormal}), has already indicated that such systems risk a recurrence, in which, after an initial, short-term, remission, the population reemerges and resumes growth. This analysis, however, only followed $\N_A(t)$ and, hence, ignored the potential stochastic effects, under which a finite population may reach $N_t = 0$ prior to its revival at large $t$. Therefore, to complete the discussion we now use our full formalism - based on $\N_A(t),\V_A(t)$ and $\Q_A(t)$ - to predict the conditions for successful treatment under these potentially recurrent dynamics.

\begin{figure}[h!]
\centering
\includegraphics[width=0.8\textwidth]{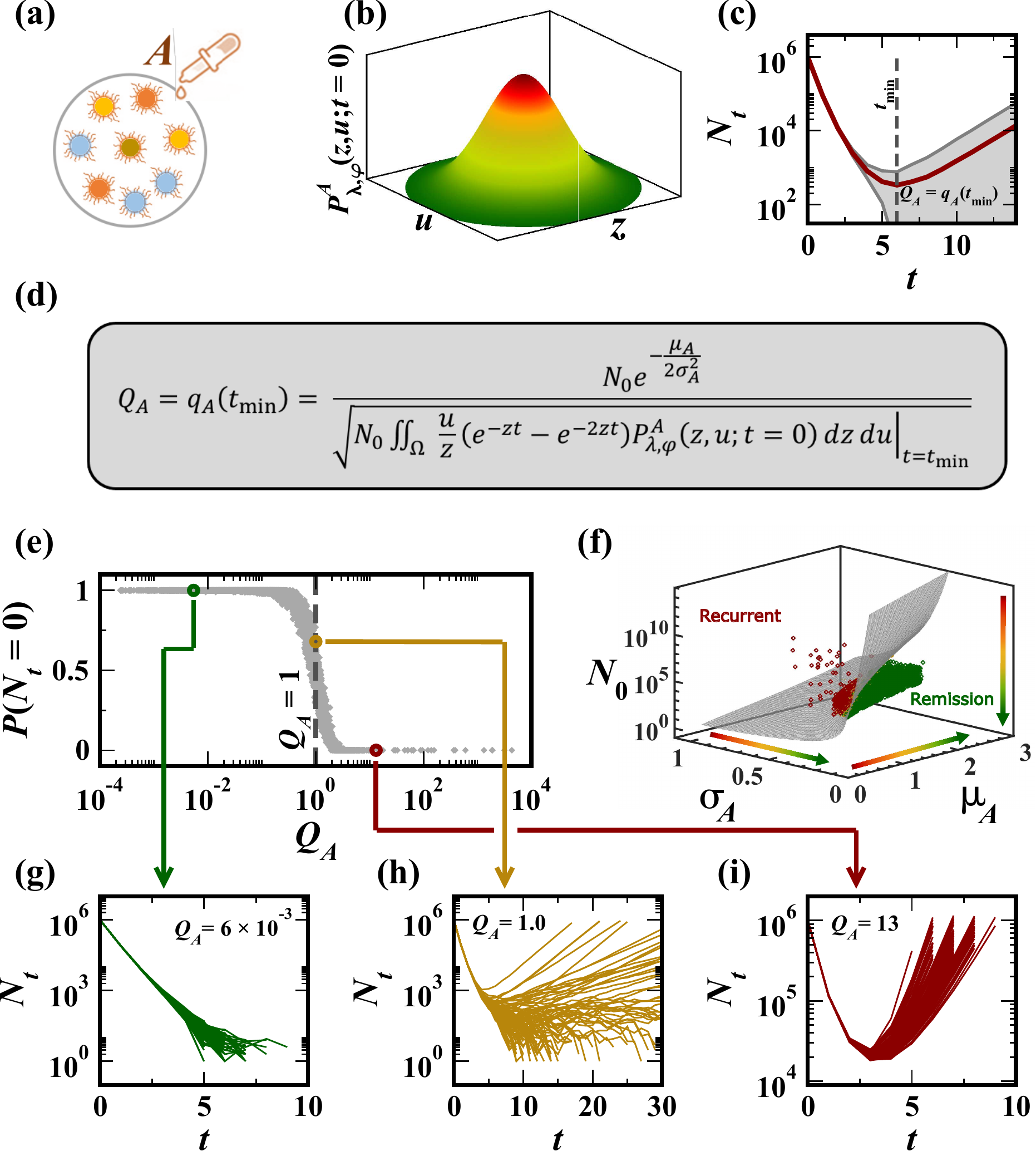}
\vspace{-3mm}
\caption{\footnotesize \textbf{\color{blue} Recurrent dynamics}.\
(a) - (b) We consider normally distributed rates in which $\lambda,\varphi \sim \Norm(\mu_A,\sigma_A^2,\nu_A,\delta_A^2)$. The distribution for $\varphi$ is truncated at zero, to ensure positive rates (see Supplementary Section 6.2). 
(c) While in this class we predict that $\N_A(t)$ will reemerge, the actual stochastic realizations $N_t$, namely the \textit{observed} dynamics, may, under some conditions, avoid recurrence and reach extinction. This is expected, if, during the population dip, when $N_t$ is at its minimum point ($t = t_{\rm min}$), a large enough fluctuation drives it towards $N_t = 0$. Hence we seek $Q_A = \Q_A(t = t_{\rm min})$, \textit{i.e}.\ the value of our statistic $\Q_A(t)$ at the population's most vulnerable time point.
(d) In Supplementary Section 4 we derive $Q_A = \Q_A(t = t_{\rm min})$, allowing us to predict the success/recurrence probability.
(e) To test the condition presented in panel (c), we examined an ensemble of $6,400$ different stochastic treatment scenarios spanning a broad range of parameter sets ($N_0,\mu_A,\sigma_A,\nu_A,\delta_A$). Each scenario was implemented across $M = 100$ independent realizations, allowing us to extract the success probability $P(N_t = 0)$ from the fraction of realizations that ended in population extinction. For each scenario we also calculated $Q_A$ as shown in panel (d). Plotting $P(N_t = 0)$ vs.\ $Q_A$ (grey dots) we find that indeed $Q_A \le 1$ (vertical dashed line) sharply splits between the recurrent scenarios ($P(N_t = 0) \to 0$, right) and the successful ones ($P(N_t = 0) \to 1$, left). Hence, our formalism allows us to evaluate \textit{a priori}, through $Q_A$, the potential efficacy of a given treatment, and its probability for recurrence. This prediction is encapsulated within the single parameter $Q_A$ in (\ref{Qmin}), as determined by the malignant population size $N_0$ and by its distributed response $P^A_{\lambda,\varphi}(z,u;t = 0)$ to the proposed treatment.  
(f) The recurrent vs.\ successful phase boundary in the $N_0,\mu_A,\sigma_A$ phase space (grey surface). The predicted boundary sharply splits between our successful (above $90\%$ remission; green) and failed (above $90\%$ recurrence, red) simulated scenarios. The arrows point in the direction of the ideal treatment:\ large average response (high $\mu_A$) and a low response variability across the cellular population (small $\sigma_A$). These considerations play an exceeding role as the initial population size $N_0$ is increased, and therefore under small $N_0$ the conditions for success are enhanced.
(g) - (i) We focused on three specific scenarios from our ensemble:\ $Q_A = 5 \times 10^{-3}$ (green), $Q_A = 1$ (orange) and $Q_A = 15$ (red). For each of these scenarios we show the $M = 100$ realizations of $N_t$ vs.\ $t$. As expected, each realization follows a distinct sequence, a testament to the random nature of $N_t$. The crucial point is that under $Q_A \ll 1$ all realizations end successfully in $N_t = 0$ (green), while for $Q_A \gg 1$ they all exhibit the predicted recurrence (red). In between, under $Q_A = 1$, we observe borderline behavior (orange), with some realizations ($\sim 40\%$) ending in recurrence, while others ($\sim 60\%$) terminating in the desired remission. 
}
\label{Fig5}
\end{figure}

The critical junction in $N_t$'s trajectory on which we focus is at the minimum point of $\N_A(t)$ in (\ref{NtNormal}), which occurs at $t_{\rm min} = \mu_A / \sigma^2_A$ (Fig.\ \ref{Fig5}c, vertical dashed line). It is at this point, where $N_t$ is closest to extinction, and can either hit zero or pull through and enter the proliferation phase (see discussion in Supplementary Section 4.1). Using condition (\ref{Qmin}) this all boils down to whether $Q_A = \Q_A(t = t_{\rm min})$ is greater or smaller than unity. Therefore in Supplementary Section 4 we derive $Q_A$ as a function of $N_0$ and $P^A_{\lambda,\varphi}(z,u;t = 0)$, obtaining an implicit integral expression that can be extracted computationally (Fig.\ \ref{Fig5}d). Next, we simulated $6,400$ scenarios with varying parameters, and examined, for each parameter set, the fraction of recurrent vs.\ extinction events. This helps us directly evaluate the probability $P(N_t = 0)$ of treatment success. In Fig.\ \ref{Fig5}e we present $P(N_t = 0)$ vs.\ $Q_A$ and observe precisely the predicted sharp transition, from a high likelihood for recurrence under $Q_A > 1$, to an almost guaranteed success when $Q_A \le 1$. 

To illustrate the different scenarios we focus, in Fig.\ \ref{Fig5}g-i, on three specific instances:\ a predictably successful treatment ($Q_A = 5 \times 10^{-3}$, green); a predictably unsuccessful treatment ($Q_A = 15$, red); and a borderline treatment, in which $Q_A = 1$ precisely (yellow). Indeed, under the green scenario all $M = 100$ stochastic realizations terminate in population extinction - the treatment is practically guaranteed to eliminate the undesired cell population. In contrast, the red scenario is bound for failure, with every realization ending in eventual recurrence. Finally, yellow, situated precisely at the boarder between success and failure, shows a divided outcome, with $\sim 40\%$ remission and $\sim 60\%$ recurrence.

Once again, we observe the crucial role of population heterogeneity, this time not in determining efficiency, as in the case of slow remission, but rather in predicting success vs.\ failure. Therefore, when assessing the potential efficacy of different treatments, they should be ranked not just based on the population average ($\mu_A$), but rather on $Q_A$, as appears in Fig.\ \ref{Fig5}d. We illustrate this in Fig.\ \ref{Fig5}f by partitioning all $N_0,\mu_A,\sigma_A$ combinations along the $Q_A = 1$ plane (grey), designed to divide treatment space between success (green) and failure (red). This points to the most favorable conditions for achieving remission (red-green arrows), namely:\ Small initial population (low $N_0$), large average decay rate (high $\mu_A$) and low cellular variability (small $\sigma_A$).

\vspace{2mm}
\textbf{\Large \color{blue} Discussion and outlook}

Matching treatment to malignancy, a clinical challenge, maps quite naturally into a population dynamics problem. In a deterministic setting, it all condenses into a single parameter $\lambda$, characterizing how rapidly the treatment eliminates the malignant cells. Here, however, we have shown that stochastic effects play a crucial role in the actual performance of a selected treatment. Specifically, we discussed two sources of randomness:\ the ingrained heterogeneity in the cells' response to the treatment, and the random nature of the sequence of events following treatment. Together, we found, that these two, practically inevitable features of real-world population dynamics, lead to a sharp, qualitative, departure from the classic deterministic solution. 

Fortunately, we show that even under this complex, seemingly unpredictable, setting the treatment outcome \textit{can} be predicted. It requires not just a single observable $\lambda$, but rather a set of probabilistic parameters:\ $\N_A(t)$ and $\V_A(t)$ of Eqs.\ (\ref{NtCumulant}) and (\ref{Variance}), and their subsequent $Q_A$ and $T_A$ in (\ref{Qmin}) and (\ref{T}). These analytically tractable parameters, rooted in the population's heterogeneous response to each treatment, provide the desired predictors of the treatment's potential success and efficiency. 

The reduction we offer here, into a limited set of quantifiable parameters, enables making systematic clinical decisions, helping match treatments to malignancies. To apply our methodology, however, we must embark on systematic experiments to measure not just the \textit{average} performance of different therapeutics, but rather their statistical response patterns $P^A_\lambda(z;t = 0)$, and from that extract their characteristic $Q_A$ and $T_A$. A first step was already undertaken in the recent experiments of Fig.\ \ref{Fig3}e-g, where researchers mapped the distributed response of cancerous cultures to chemotherapeutic agents \cite{Jacob2021} (see Supplementary Section 6.3). Expanding this line of experimentation to other malignancies/treatments, will help clinicians make informed, mathematically driven, treatment prioritization.      
     
\clearpage

\clearpage

\clearpage

\clearpage

\clearpage

\clearpage

\bibliographystyle{unsrt}
\bibliography{../bib/bib}
\clearpage

\centering

\textbf{\Large \color{blue} Figures - Full size}

\begin{figure}[h!]
\centering
\includegraphics[width=1\textwidth]{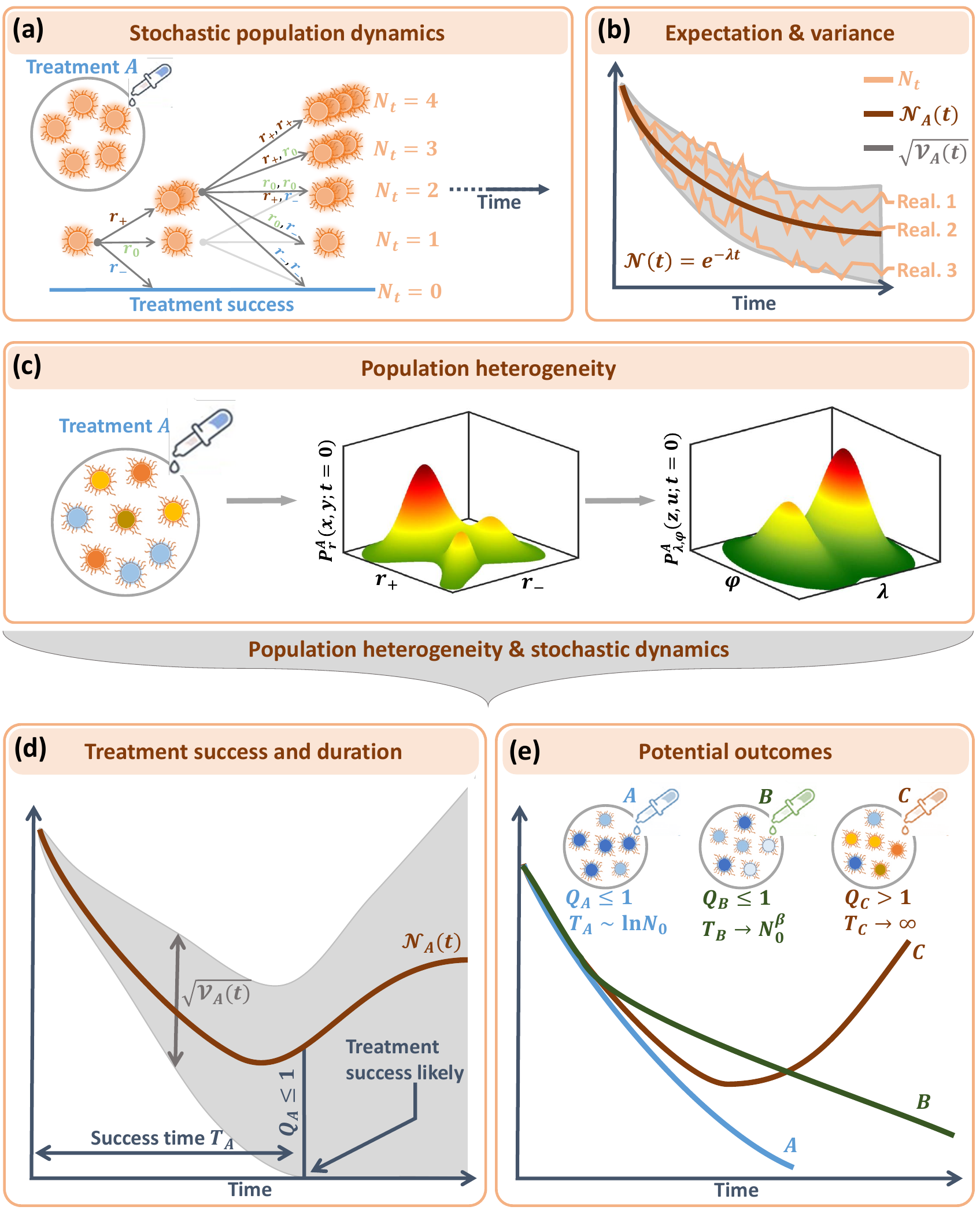}
\end{figure}
\textbf{\color{blue} Figure 1.\ The malignancy-treatment interplay}.\

\clearpage

\begin{figure}[h!]
\centering
\includegraphics[width=1\textwidth]{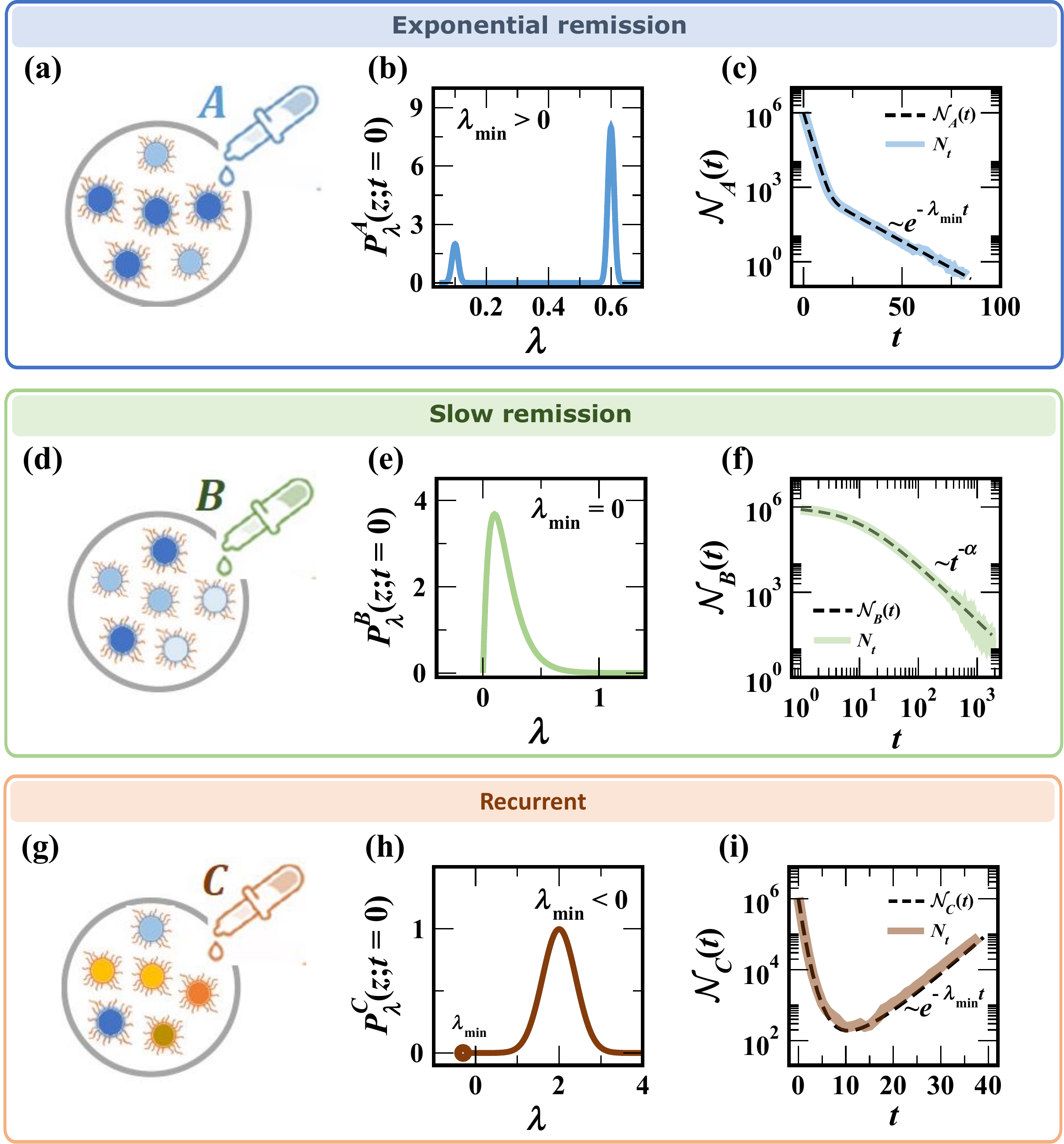}
\end{figure}
\textbf{\color{blue} Three classes of the \textit{expected} population dynamics $\N_A(t)$}.\

\clearpage

\begin{figure}[h!]
\centering
\includegraphics[width=0.94\textwidth]{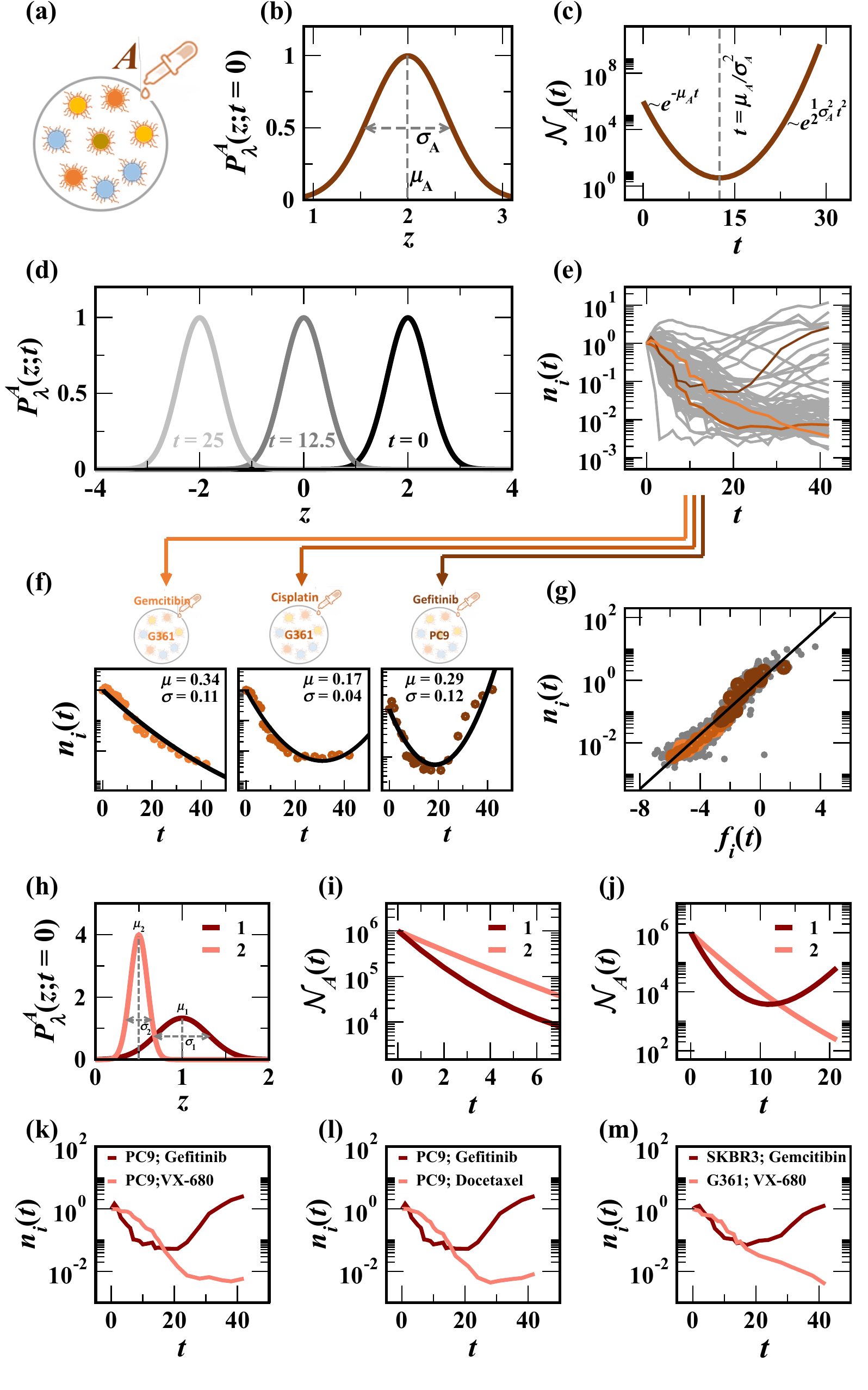}
\end{figure}
\vspace{-7mm}
\textbf{\color{blue} Expected trajectory $\N_A(t)$ under normally distributed rates}.\

\clearpage

\begin{figure}[h!]
\centering
\includegraphics[width=1\textwidth]{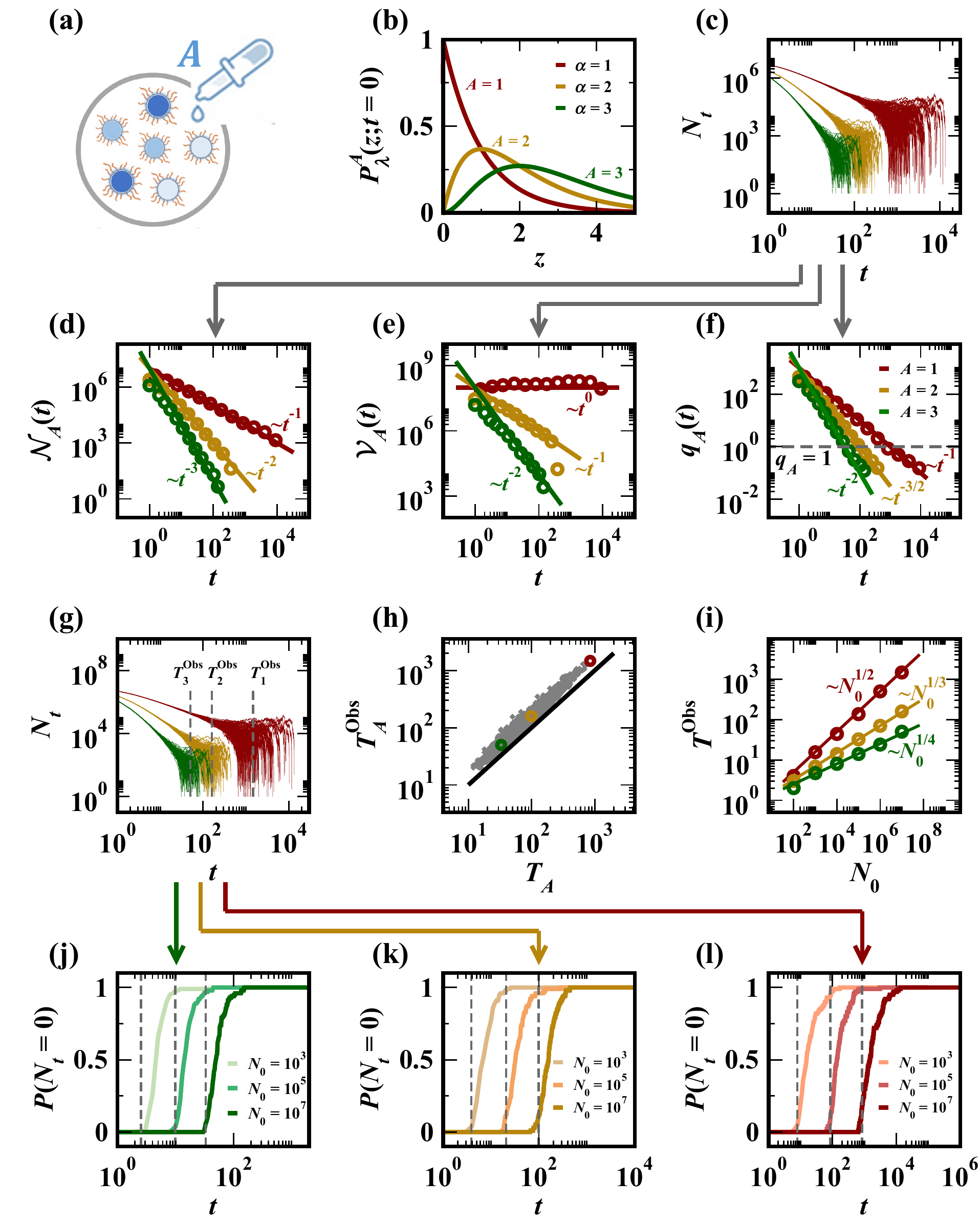}
\end{figure}
\vspace{-7mm}
\textbf{\color{blue} Slow remission}.\

\clearpage

\begin{figure}[h!]
\centering
\includegraphics[width=1\textwidth]{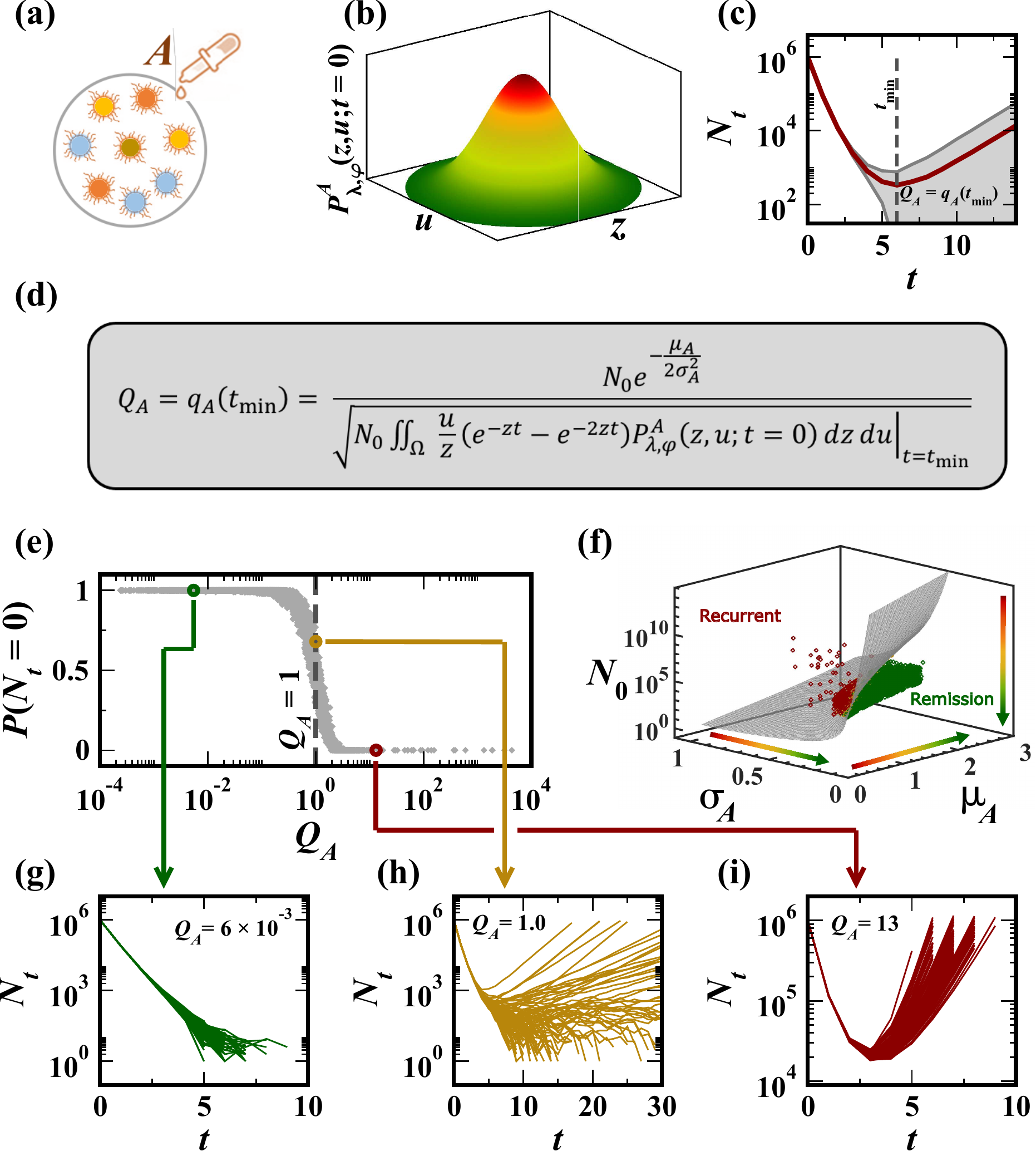}
\end{figure}
\vspace{-5mm}
\textbf{\color{blue} Recurrent dynamics}.\

\end{document}